\documentclass[aps,pra,twocolumn,superscriptaddress,longbibliography,showpacs, 10pt]{revtex4-1}
\usepackage[english]{babel}
\usepackage{graphicx}
\usepackage{color}
\usepackage{float}
\usepackage{siunitx}

\begin{document}

\newcommand{\fceo}{$f_{\mathrm{CEO}}$ }
\newcommand{\about}{$\sim$}
\newcommand{\sio}{$\mathrm{SiO_2}$ }

\title{Ultrabroadband supercontinuum generation and frequency-comb stabilization using on-chip waveguides with both cubic and quadratic nonlinearities}

\newcommand{\NISTTF}{Time and Frequency Division, National Institute of Standards and Technology, Boulder, Colorado 80305, U.S.A.}
\newcommand{\NISTAP}{Applied Physics Division, National Institute of Standards and Technology, Boulder, Colorado 80305, U.S.A.}
\newcommand{\NISTG}{Center for Nanoscale Science and Technology, National Institute of Standards and Technology, Gaithersburg, Maryland 20899, U.S.A.}
\newcommand{\YALE}{Department of Electrical Engineering, Yale University, New Haven, Connecticut, 06520, U.S.A}
\newcommand{\CU}{Department of Physics, University of Colorado, Boulder, Colorado, 80309, U.S.A.}

\author{Daniel~D.~Hickstein}  \email[]{danhickstein@gmail.com} \affiliation{\NISTTF}
\author{Hojoong~Jung}         \affiliation{\YALE  }
\author{David~R.~Carlson}     \affiliation{\NISTTF}
\author{Alex~Lind}            \affiliation{\NISTTF} \affiliation{\CU}
\author{Ian~Coddington}       \affiliation{\NISTAP}
\author{Kartik~Srinivasan}    \affiliation{\NISTG }
\author{Gabriel~G.~Ycas}      \affiliation{\NISTTF}
\author{Daniel~C.~Cole}       \affiliation{\NISTTF} \affiliation{\CU}
\author{Abijith~Kowligy}      \affiliation{\NISTTF}
\author{Connor~Fredrick}      \affiliation{\NISTTF} \affiliation{\CU}
\author{Stefan~Droste}        \affiliation{\NISTAP}
\author{Erin~S.~Lamb}         \affiliation{\NISTTF}
\author{Nathan~R.~Newbury}    \affiliation{\NISTAP} 
\author{Hong~X.~Tang}         \affiliation{\YALE  } 
\author{Scott~A.~Diddams}     \affiliation{\NISTTF} \affiliation{\CU} 
\author{Scott~B.~Papp}        \affiliation{\NISTTF}

\date{\today}

\begin{abstract}
Using aluminum-nitride photonic-chip waveguides, we generate optical-frequency-comb supercontinuum spanning from 500~nm to 4000~nm with a 0.8 nJ seed pulse, and show that the spectrum can be tailored by changing the waveguide geometry. Since aluminum nitride exhibits both quadratic and cubic nonlinearities, the spectra feature simultaneous contributions from numerous nonlinear mechanisms: supercontinuum generation, difference-frequency generation, second-harmonic generation, and third-harmonic generation. As one application of integrating multiple nonlinear processes, we measure and stabilize the carrier-envelope-offset frequency of a laser comb by direct photodetection of the output light. Additionally, we generate $\sim$0.3 mW in the 3000~nm to 4000~nm region, which is potentially useful for molecular spectroscopy. The combination of broadband light generation from the visible through the mid-infrared, combined with simplified self-referencing, provides a path towards robust comb systems for spectroscopy and metrology in the field.
\end{abstract}

\maketitle

\section{Introduction \label{intro}}

Optical frequency combs are laser-based light sources that enable a wide variety of precision measurements, including the comparison of state-of-the-art atomic clocks~\cite{rosenband_frequency_2008}, the quantitative measurement of pollution over several-kilometer paths above cities~\cite{rieker_frequency-comb-based_2014, waxman_intercomparison_2017}, and even the search for distant Earth-like planets \cite{li_laser_2008, ycas_demonstration_2012}. Laser frequency combs are typically generated with relatively narrow (\about10 \%) relative spectral bandwidth \cite{kippenberg_microresonator-based_2011}. However, broad bandwidth is a requirement for many applications, such as spectroscopy, where it is desirable to probe several atomic or molecular transitions simultaneously, and optical frequency metrology, where stable lasers at different wavelengths must be compared. Consequently, narrowband frequency combs are usually spectrally broadened to at least one octave via supercontinuum generation (SCG) in materials with cubic nonlinearity ($\chi^{(3)}$), such as highly nonlinear fiber (HNLF) or photonic crystal fiber \cite{dudley_supercontinuum_2006}. 

\begin{figure}[h!]
	\includegraphics[width=\linewidth]{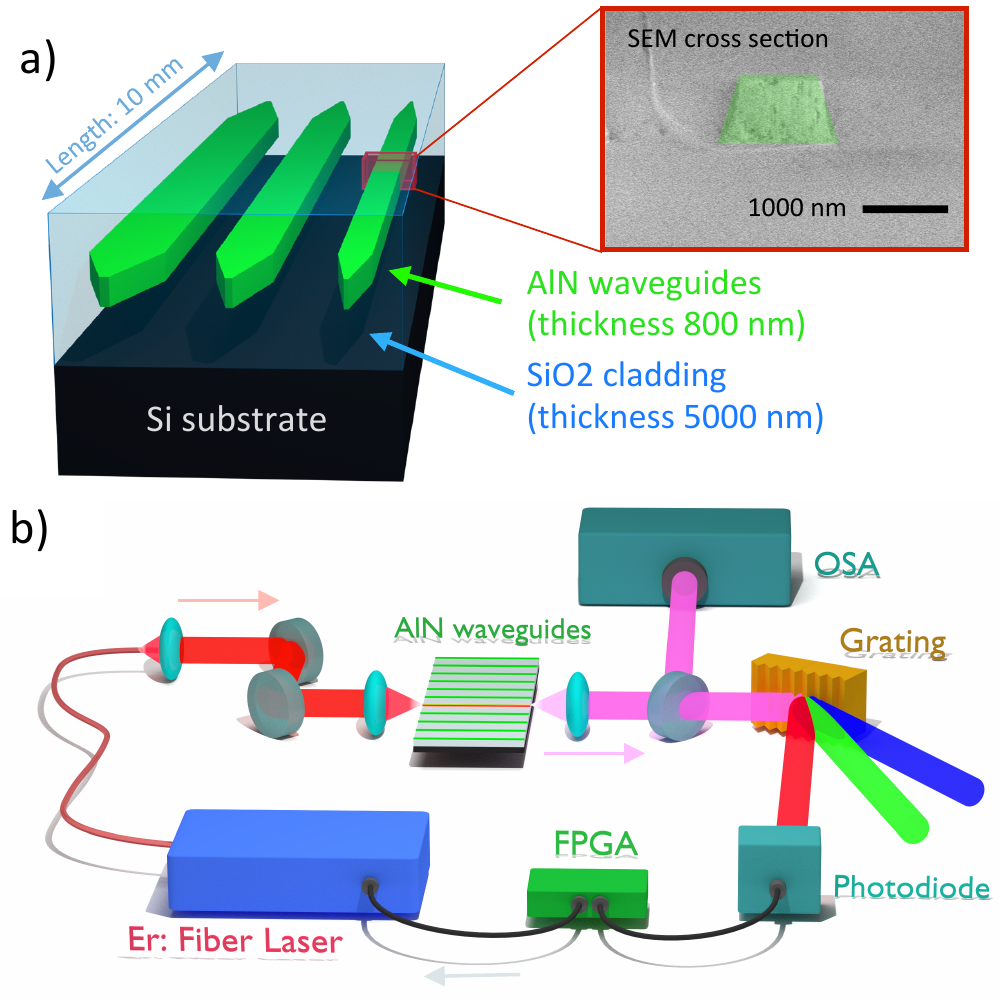}
	\caption{\label{fig:overview} a) Aluminum nitride (AlN) on-chip waveguides embedded in \sio  tightly confine the light-field, providing high nonlinearity. b) To generate supercontinuum, 80-fs laser pulses (1560 nm, 800 pJ) are coupled into each waveguide. The broadband output is directed into an optical spectrum analyzer (OSA), or dispersed with a grating, where \fceo is detected in the 780-nm region using a photodiode. The \fceo signal is digitized using a field-programmable gate-array (FPGA), which applies feedback to the laser pump diode.}
\end{figure}

Moreover, octave-spanning bandwidth allows the carrier-envelope-offset frequency ($f_{\mathrm{CEO}}$) of the frequency comb to be measured (and subsequently stabilized) using ``f--2f'' self referencing \cite{jones_carrier-envelope_2000, holzwarth_optical_2000,diddams_direct_2000}. In the f--2f scheme, the low frequency portion of the spectrum undergoes second harmonic generation (SHG) in a material with quadratic nonlinearity ($\chi^{(2)}$), such as $\mathrm{LiNbO_3}$, and interferes with the high-frequency portion of the spectrum, producing a signal that oscillates at $f_{\mathrm{CEO}}$. Due to the modest effective nonlinearity of silica HNLF, SCG using traditional silica fiber requires high peak powers (typically 10 kW or more), which increases the electrical power requirements of the laser and limits the achievable repetition rates. Indeed, the adoption of new and compact frequency comb sources at gigahertz repetition rates, such as electro-optic combs \cite{kobayashi_highrepetitionrate_1972,torres-company_optical_2014} and microresonator combs \cite{kippenberg_microresonator-based_2011, delhaye_optical_2007, herr_temporal_2014}, is currently hindered by the difficulty of generating octave-spanning spectra using low-peak-power pulses. In addition, many potential applications of frequency combs require supercontinuum light at wavelengths that are difficult to achieve with SCG in silica fiber. In particular, light in the mid-infrared (\SI{3}{\micro\meter} to \SI{8}{\micro\meter}) region is advantageous for molecular spectroscopy \cite{schliesser_mid-infrared_2012, coddington_dual-comb_2016,truong_dual-comb_2016, giorgetta_broadband_2015, cossel_gas-phase_2017}, but is absorbed by silica fiber.
%

Fortunately, on-chip photonic waveguides with wavelength-scale dimensions offer high confinement of light, which provides a substantial increase in the effective nonlinearity 
\begin{equation}
\label{eq:gamma}
\gamma = \frac{2 \pi n_2}{\lambda A_\mathrm{eff}},
\end{equation}
where $\lambda$ is the wavelength, $A_{\mathrm{eff}}$ is the effective area of the mode, and $n_2$ is the material-dependent nonlinear index, which is directly proportional to $\chi^{(3)}$ \cite{dudley_supercontinuum_2006}. In addition, materials with higher $\chi^{(3)}$ -- such as silicon nitride~\cite{epping_chip_2015, porcel_two-octave_2017, klenner_gigahertz_2016, mayer_frequency_2015, boggio_dispersion_2014, hickstein_photonic-chip_2016, carlson_photonic-chip_2017, johnson_octave-spanning_2015}, silicon~\cite{singh_midinfrared_2015, hsieh_supercontinuum_2007, leo_coherent_2015}, aluminum gallium arsenide~\cite{pu_supercontinuum_2016}, and chalcogenide materials \cite{yu_mid-infrared_2013,lamont_supercontinuum_2008} -- further increase $\gamma$ and allow much lower peak power (${<}1$ kW) to be used for the SCG process. High confinement waveguides provide the additional advantage of increased control over the group-velocity dispersion (GVD), and therefore the spectral output of the SCG process. 

Currently, supercontinuum generation in materials with both strong $\chi^{(2)}$ and $\chi^{(3)}$ nonlinearities is opening new possibilities for broadband light sources. For example, experiments with periodically poled $\mathrm{LiNbO_3}$ (PPLN) have demonstrated supercontinuum generation via cascaded $\chi^{(2)}$ processes, and the simultaneous generation of supercontinuum and harmonic light \cite{iwakuni_generation_2016, guo_supercontinuum_2015, langrock_generation_2007}. Recently, aluminum nitride (AlN) has emerged as a lithographically compatible material that exhibits both strong $\chi^{(2)}$ and $\chi^{(3)}$ nonlinearities in addition to a broad transparency window. Consequently, thin-film AlN is proving to be a versatile platform for nanophotonics, providing phase-matched second-harmonic generation (SHG) \cite{guo_second-harmonic_2016}, frequency comb generation \cite{jung_optical_2013}, and ultraviolet light emission \cite{zhao_aluminum_2015}.  

Here we present the first observations of SCG in lithographically fabricated, on-chip AlN waveguides and demonstrate that the platform provides exciting new capabilities: (1) We observe SCG from 500~nm to 4000~nm, and show the spectrum can be tailored simply by changing the geometry of the waveguide. (2) We find that the material birefringence induces a crossing of the transverse-electric (TE) and transverse-magnetic (TM) modes, which enhances the spectral brightness in a narrow band, and that the spectral location of this band can be adjusted by changing the waveguide dimensions. (3) We observe bright SHG, which is phase-matched via higher-order modes of the waveguide, as well as phase-mismatched difference frequency generation (DFG), which produces broadband light in the 3500~nm to 5500~nm region. (4) We demonstrate that simultaneous SCG and SHG processes in an AlN waveguide allows \fceo to be extracted directly from the photodetected output, with no need for an external SHG crystal, recombination optics, or delay stage. (5) We use this simple scheme to lock the \fceo  of a compact laser frequency comb, and find that the stability of the locked \fceo is comparable to a standard $f$--$2f$ interferometer and sufficient to support precision measurements.

\section{Experiment\label{experiment}}
%
The fully $\mathrm{SiO_2}$-clad AlN waveguides~\cite{jung_optical_2013, xiong_low-loss_2012} have a thickness (height) of 800~nm, and a width that varies from 400~nm to 5100~nm. Near the entrance and exit facets of the chip, the waveguide width tapers to 150~nm in order to expand the mode and improve the coupling efficiency, which is estimated at -4~dB/facet, on average. We generate supercontinuum by coupling into the waveguide approximately 80~mW of 1560~nm light from a compact, turn-key Er-fiber frequency comb \cite{sinclair_compact_2015}, which produces $\sim$80~fs pulses at 100~MHz. The polarization of the light is controlled using achromatic quarter- and half-waveplates. The light is coupled into each waveguide using an aspheric lens (NA=0.6) designed for 1550 nm. For output coupling, two different techniques are used, as shown in Fig.~\ref{fig:overview}b. In the case of \fceo detection, the light is out-coupled using a visible wavelength microscope objective (NA=0.85) and then dispersed with a grating before illuminating a photodiode.  Alternatively, when recording the spectrum, the light is collected by butt-coupling a $\mathrm{InF_3}$ multimode fiber (NA=0.26) at the exit facet of the chip. The waveguide output is then recorded using two optical spectrum analyzers (OSAs); a grating-based OSA is used to record the spectrum across the visible and near-infrared regions, while a Fourier-transform OSA extends the coverage to 5500~nm.

\begin{figure}
	\includegraphics[width=\linewidth]{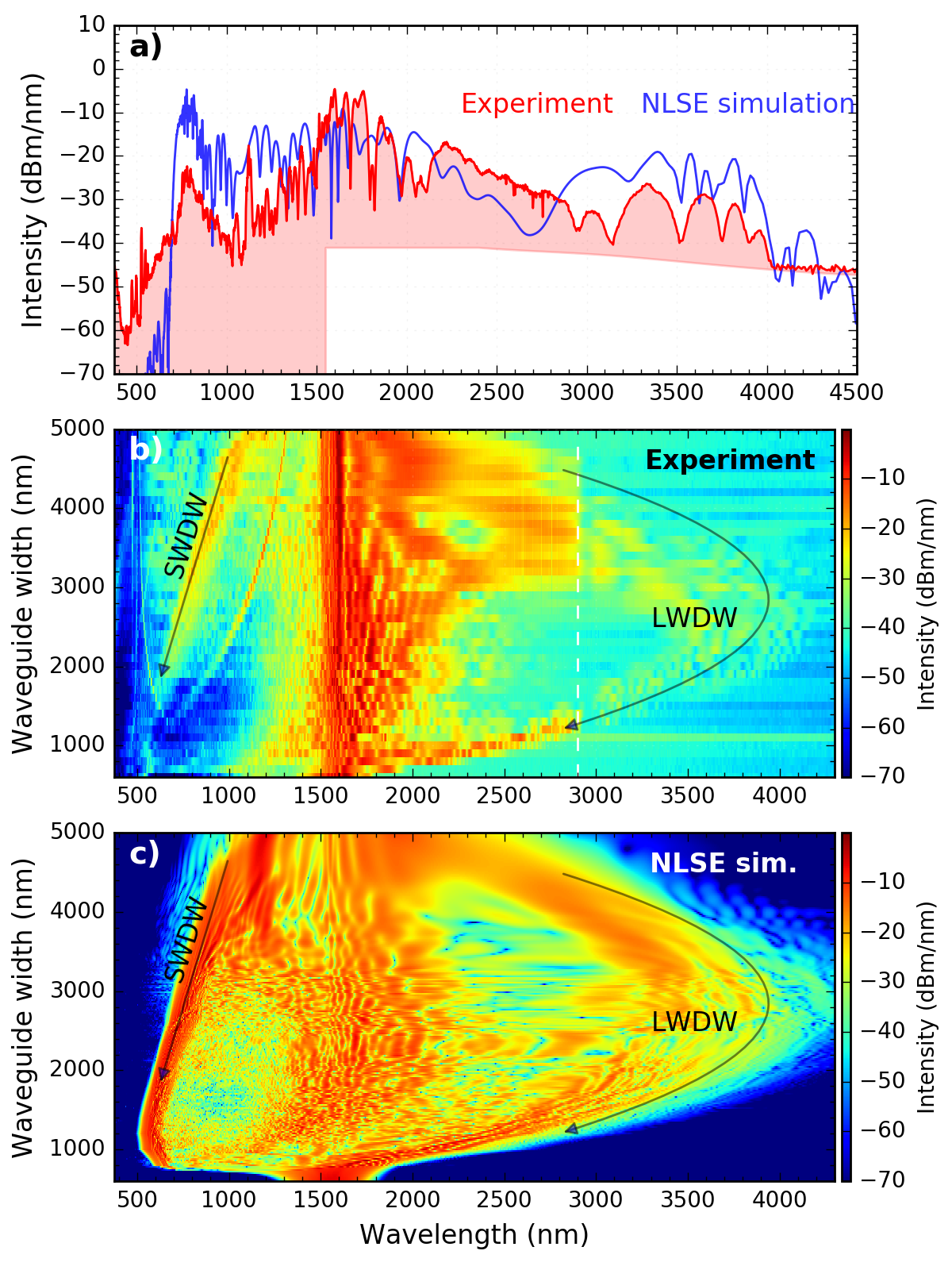}
 	\caption{\label{fig:TE} Supercontinuum generation from the lowest order quasi-transverse-electric ($\mathrm{TE_{00}}$) mode. a) Experimental and theoretical optical spectrum from the 3200-nm wide wave\-guide (scaled by +7 dB to compensate for output-coupling to multimode fiber). The bottom of the shaded region indicates the noise-floor of the OSA. b) Experimentally observed spectra from all waveguide widths on the chip. The dashed line at 2900~nm indicates the onset of long-wavelength absorption in the waveguides. c) Simulated spectra using the Nonlinear Schr\"odinger Equation (NLSE) are in general agreement with experiment, and suggest that wavelength-dependent absorption is decreasing the amount of mid-infrared light observed experimentally. Solid lines indicate the short-wavelength and long-wavelength dispersive waves (SWDW and LWDW), and are in the same location in both (b) and (c).}
\end{figure}

To model the supercontinuum generation, we perform numerical simulations using the nonlinear Schr\"odinger equation (NLSE), as implemented in the PyNLO package \cite{hult_fourth-order_2007, heidt_efficient_2009, ycas_pynlo_2016, amorim_sub_2009}. The effective refractive indices and effective nonlinearities of the waveguides are calculated using the vector finite-difference modesolver of Fallahkhair, Li, and Murphy \cite{fallahkhair_vector_2008}. The NLSE includes $\chi^{(3)}$ effects and incorporates the full wavelength dependence of the effective index, but it does not take into account any $\chi^{(2)}$ effects, higher order modes, or wavelength-dependent absorption.


\section{Results and Discussion}

\subsection{Supercontinuum from visible to mid-infrared}
When pumped in the lowest-order quasi-transverse-electric mode ($\mathrm{TE_{00}}$), the AlN waveguides generate light (Fig.~\ref{fig:TE}) from the blue portion of the visible region (\about500~nm) to the mid-infrared (\about4000~nm). The broad peaks on both sides of the spectrum are the short-wavelength and long-wavelength dispersive waves (labeled ``SWDW'' and ``LWDW'' in Fig. \ref{fig:TE}b,c), which are generated at locations determined by the GVD of the waveguide \cite{akhmediev_cherenkov_1995, dudley_supercontinuum_2006}. The broadband spectrum is a result of the flat GVD profile enabled by strong confinement of the light in these waveguides. The simulated spectra (Fig.~\ref{fig:TE}c) reproduce the spectral location of thee long-wavelength and short-wavelength dispersive waves. However, the NLSE simulations overestimate the light intensity in the dispersive waves compared with the experiment. One reason for this discrepancy is that the waveguide mode at 1560~nm does not have perfect overlap with modes at different wavelengths, and the effective nonlinearity is actually  smaller than what is predicted by Eq. \ref{eq:gamma}, which assumes perfect mode-overlap. This effect is most pronounced at longer wavelengths, where the mode extends significantly outside of the waveguide and does not overlap well with the 1560~nm mode, which is mostly confined within the AlN waveguide.

When waveguide widths near 3500~nm are used, the supercontinuum shows high spectral intensity over a broad region from 1400~nm to 2800~nm, generally remaining within $-20$~dB of the transmitted pump intensity. This bright spectrum represents a promising source for molecular spectroscopy, since OH stretching transitions absorb in this region \cite{solomons_organic_2009}. Indeed, sharp dips visible in the spectral intensity near 2700~nm are due to the absorption of water vapor in the OSA. Unfortunately, a sharp minimum in the spectrum near 2900~nm, and decreased intensity at wavelengths longer that 2900~nm suggests that these mid-infrared wavelengths are not efficiently transmitted through the waveguides. This loss is likely due to OH absorption \cite{navarra_oh_2005} in the $\mathrm{SiO_2}$, since a significant fraction of the mode extends outside the AlN waveguide and into the \sio cladding at these wavelengths. In the future, the use of a different cladding material could increase the output of mid-infrared light. Nevertheless, the waveguides still produce usable, broadband light in the mid-infrared region -- for example, we estimate that the 2600-nm waveguide produces ${\sim}0.3$ mW in the 3500~nm to 4000~nm spectral region, which is sufficient power for some applications. Indeed, the mid-infrared light is easily seen in Fig. \ref{fig:TE}b, which presents spectra collected with just a few seconds integration time for each spectrum.

\subsection{Brightness enhancement via a mode crossing}
\begin{figure}
	\includegraphics[width=\linewidth]{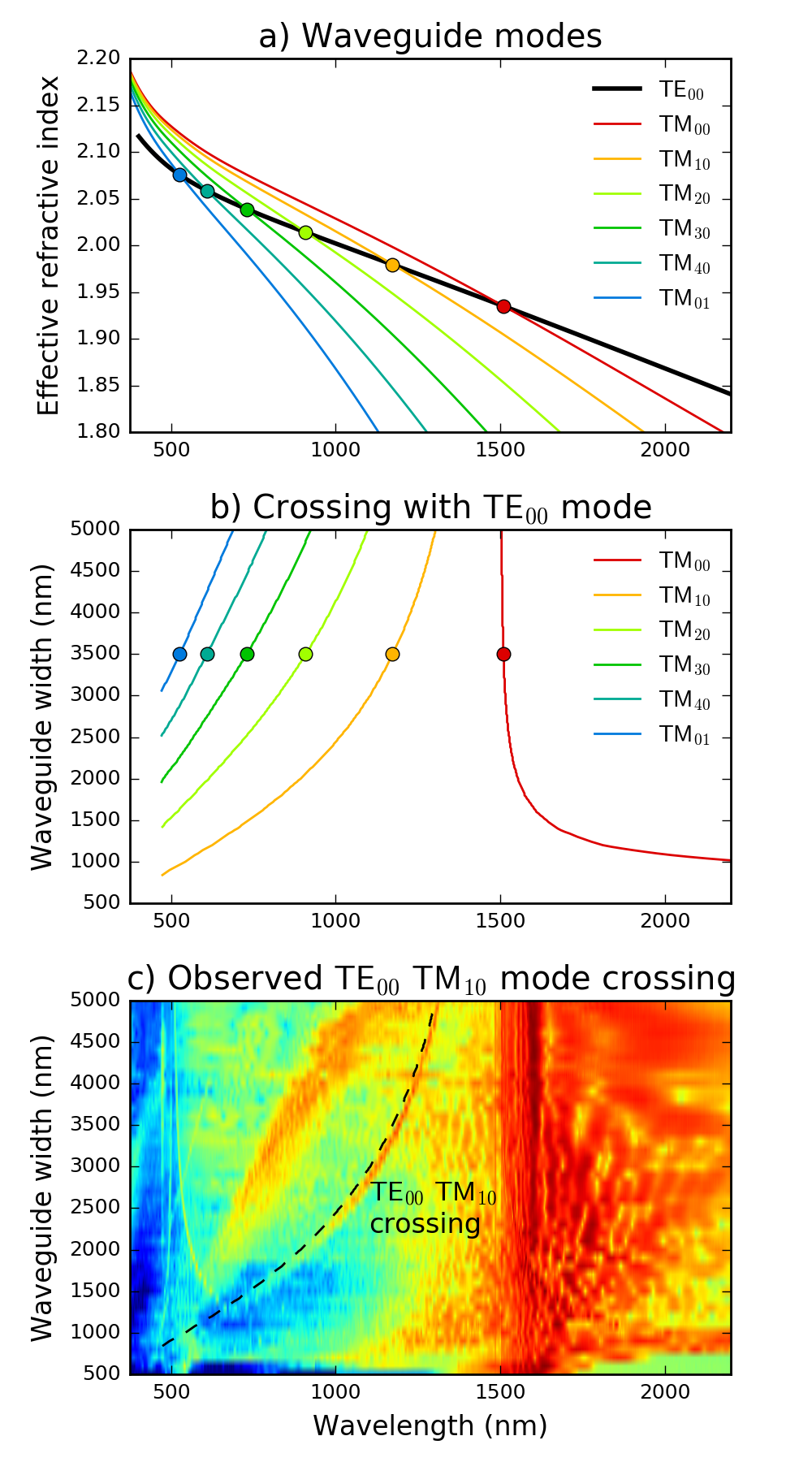}	
    \caption{\label{fig:modeCrossing} a) As the wavelength increases, the refractive index of the fundamental TE mode ($\mathrm{TE_{00}}$) crosses several TM modes. A waveguide width of 3500~nm is shown. b) The spectral location of these polarization-mode crossings changes as a function of the waveguide width (shown) and thickness (not shown). c) The crossing of the $\mathrm{TE_{00}}$ and $\mathrm{TM_{01}}$ modes (as calculated from only the bulk refractive index and waveguide geometry) matches the location of the sharp peak in the experimental spectra.}
\end{figure}

In the 800~nm to 1200~nm region, a sharp peak is seen in the supercontinuum spectrum for waveguide widths $>$1500~nm (Figs. \ref{fig:TE}b and \ref{fig:modeCrossing}c), which is not explained by the NLSE. The location of the peak occurs at the wavelength where the refractive index of the lowest order TE mode ($\mathrm{TE_{00}}$) and a higher order quasi-TM mode ($\mathrm{TM_{10}}$) cross (Fig.~\ref{fig:modeCrossing}a). While such mode crossings are commonplace in Kerr-comb generation in microring resonators \cite{cole_soliton_2016, ramelow_strong_2014, herr_mode_2014}, they are not typically seen in supercontinuum generation in straight waveguides, because the $\mathrm{TE_{00}}$ usually has the highest effective index at all wavelengths. In the case of AlN waveguides, the polarization-mode crossing occurs because AlN is a birefringent material, and the bulk index for the vertical (TM) polarization is higher than for the horizontal (TE) polarization. At short wavelengths, where the waveguide geometry provides only a small modification to the refractive index, the TM modes tend to have the highest effective index. However, at longer wavelengths, geometric dispersion plays a larger role, lowering the effective index of the TM modes more than the TE modes and causing the polarization-mode crossing. Similarly, since modifications of the waveguide width tend to change the effective index of the TE modes more than the TM modes, the spectral location of the mode crossing also depends on the width of the waveguide (Fig.~\ref{fig:modeCrossing}b).

A mode crossing causes a sharp feature in the GVD, which can allow for the phase-matching of four-wave-mixing processes in spectral regions that would otherwise be phase-mismatched \cite{cole_soliton_2016, ramelow_strong_2014}. Indeed, the crossing of the $\mathrm{TE_{00}}$ and $\mathrm{TM_{10}}$ modes enables a strong enhancement of the supercontinuum spectrum in a spectral region that is otherwise dim. In some cases, this mode crossing enables a ${\sim}25$ dB enhancement of the spectral intensity. This enhancement enables a new degree of control over the spectral output, providing a narrow, bright region that could, for example, be used to measure a heterodyne beat with a narrow-band atomic-clock laser. It is not clear why the crossing with the $\mathrm{TM_{10}}$ mode is clearly seen in the experiment, while the crossings with the higher order TM modes are absent. Understanding what mechanism couples the modes, and how this coupling could be enhanced, would allow for further customization of the spectral output of this supercontinuum source.

\subsection{Second harmonic generation and difference frequency generation}
Since AlN has $\chi^{(2)}$ nonlinearity,  it is capable of three-wave mixing processes, such as difference frequency generation (DFG), sum-frequency generation (SFG), and SHG. The thin AlN films used in this study are not single crystals, but instead consist of many hexagonal columns, which have the crystal $z$-axis oriented in the same (vertical) direction \cite{xiong_low-loss_2012}, but a random orientation for the other crystal axes. Consequently, while there is a strong $\chi^{(2)}$ component in the vertical (TM) direction, the $\chi^{(2)}$ in the horizontal (TE) direction is much weaker. 

\begin{figure}[ht]
	\includegraphics[width=\linewidth]{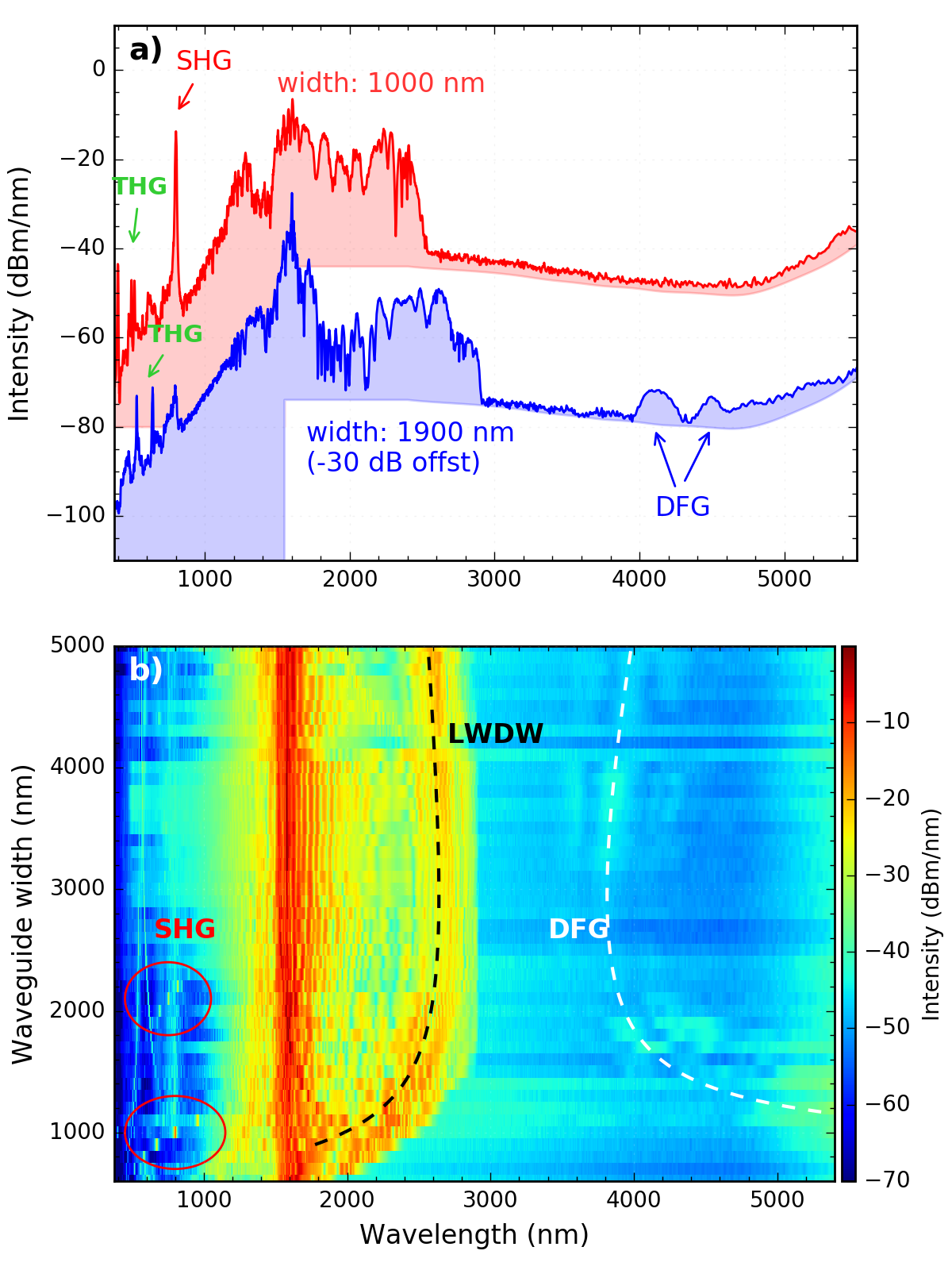}
	\caption{\label{fig:TM} Supercontinuum generation from the lowest order quasi-transverse-magnetic ($\mathrm{TM_{00}}$) mode. a) Experimental spectra from both the 1000-nm and 1700-nm width waveguides show simultaneous supercontinuum generation, second-harmonic generation (SHG), third-harmonic generation (THG), and difference-frequency generation (DFG). b) Experimental spectra from all waveguide widths, showing that waveguide geometry affects the positions of the long-wavelength dispersive wave (LWDW), the DFG peaks, and the phase-matched-SHG peaks.}
\end{figure}

Indeed, we observe the strongest  $\chi^{(2)}$ effects with the laser in the $\mathrm{TM_{00}}$ mode. The brightest SHG results from situations where the phase-velocity of the second harmonic in a higher order mode is the same as the phase velocity of the fundamental wavelength in the lowest order mode. This situation provides excellent phase matching, and we observe situations where the spectral intensity of the second harmonic light is on the same order-of-magnitude as that of the transmitted pump laser (Fig.~\ref{fig:TM}a,b). However, this phase-matching mechanism provides a phase-matching bandwidth of only a few nanometers. Additionally, we also see THG, which is phase matched to higher order modes of the waveguide.

Under TM-pumping, the waveguides also produce broadband light in the 3500~nm to 5500~nm region via DFG (Fig.~\ref{fig:TM}a,b). This process corresponds to the difference frequency between the spectrally broadened pump (1400~nm to 1700~nm) and the long-wavelength dispersive wave (2000~nm to 2700~nm). As the waveguide width becomes narrower and the dispersive wave moves to shorter wavelengths, the DFG is pushed to longer wavelengths, as determined by conservation of (photon) energy.  Indeed, for waveguide widths less than 1800 nm, the DFG moves to wavelengths longer than 5500~nm, which is outside of the range of our OSA. Additionally, the DFG process is strongly phase-mismatched, and therefore the conversion efficiency is low. However, in principle, it is possible to achieve phase matching by launching the pump laser into a higher-order mode of the waveguide.

\begin{figure*}
	\includegraphics[width=\linewidth]{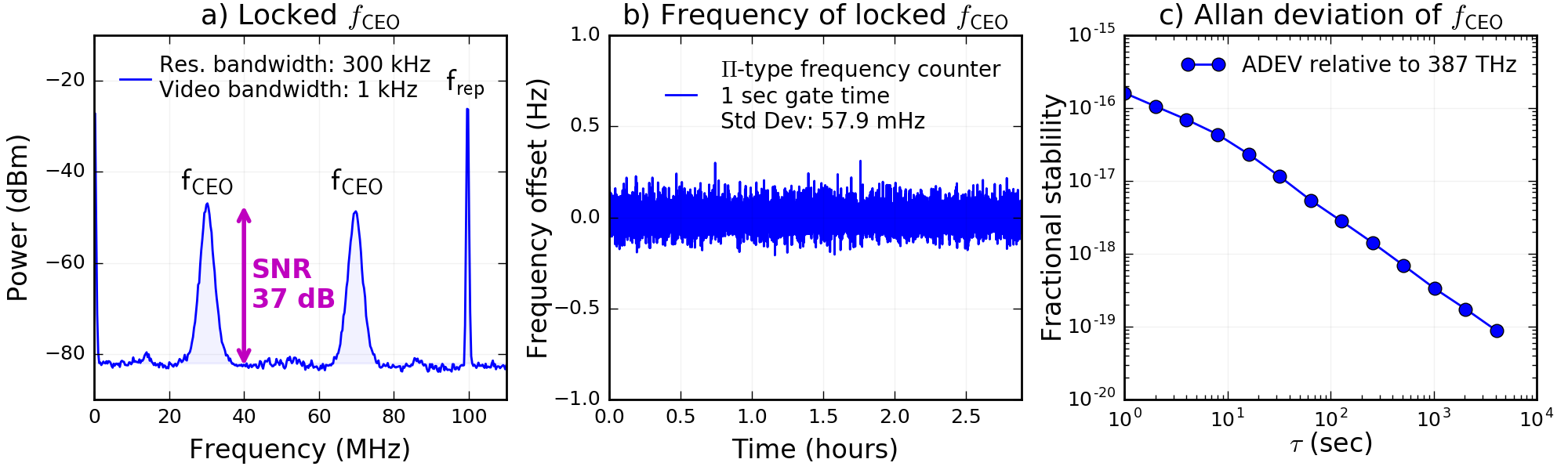}
	\caption{\label{fig:stability} a) When the \about780-nm region of the supercontinuum is detected with a photodiode, the \fceo can be observed directly, without the need for an interferometer. b) The frequency of the locked \fceo is stable over many hours. c) The (ordinary) Allan deviation of the frequencies shown in (b) demonstrates that the comb has been stabilized to a level suitable for precision metrology.}
\end{figure*}

\subsection{$\mathbf{f_{ceo}}$ detection and comb stabilization}
Since AlN exhibits both $\chi^{(2)}$ as well as strong $\chi^{(3)}$, \fceo can be directly detected in the 780-nm region, as a result of simultaneous SHG and SCG. Unlike a traditional f--2f measurement, no interferometer is needed to set the temporal overlap of the interfering beams, and no additional alignment is necessary. The only equipment required to detect \fceo is a 780-nm bandpass filter and a photodetector. Since these AlN waveguides have the strongest $\chi^{(2)}$ tensor component in the vertical direction, we observe the highest signal-to-noise ratio \fceo signal when pumping in the $\mathrm{TM_{00}}$ mode. When TM pumping the 4800-nm-width waveguide, we achieve ~37 dB SNR for the \fceo peak (Fig.~\ref{fig:stability}a). Interestingly, the highest SNR \fceo was obtained from phase-mismatched SHG in the larger width waveguides, despite the fact that much higher efficiency phase-matched SHG was seen for waveguide widths near 1000~nm. 

We speculate that the poor mode overlap between the supercontinuum (in the $\mathrm{TM_{00}}$ mode) and the phase-matched second harmonic (in a higher-order TM mode) hinders detection of the $f_{\mathrm{CEO}}$. Indeed, a recent attempt to detect a $f$--$3f$ signal in SiN waveguides found that mode overlap severely limited the achievable SNR \cite{carlson_high-efficiency_2017}. In contrast, the phase-mismatched SHG that takes place in the fundamental mode compensates for low conversion-efficiency with better overlap with the supercontinuum light. Furthermore, the highest SHG conversion likely takes place at the point of soliton fission, where the pulse is compressed and the peak intensity is the highest. This is the same point where most of the supercontinuum light is generated. Since the $f$ and $2f$ signals are generated simultaneously, and propagate in the same waveguide mode, temporal overlap is provided automatically. Nevertheless, in future implementations, on-chip mode converters \cite{guo_chip_2016} could be used to provide both phase-matched SHG, as well as mode overlap, thereby providing higher \fceo signal.

With the \fceo detected directly from the waveguide output (Fig. \ref{fig:overview}b), we could achieve glitch-free \fceo locking of a compact frequency comb for several hours (Fig.~\ref{fig:stability}b). By recording the frequency of the \fceo beat with an independent $\Pi$-type \cite{dawkins_considerations_2007} frequency counter (Fig.~\ref{fig:stability}c), we can verify that the \fceo has been stabilized to a level comparable to what can be achieved with a traditional f--2f interferometer \cite{sinclair_compact_2015}. Unfortunately, thermal drifts in the input coupling prevented locking for more than a few hours without re-alignment. In the future, input and output coupling could be accomplished via fibers glued to the facets of the chip \cite{jung_phase-dependent_2016}, which would effectively eliminate thermal drift in the coupling, and enable long-term stabilization of the laser comb.

\section{Conclusion}
In summary, we have demonstrated aluminum nitride, a lithographically compatible material with strong $\chi^{(2)}$ and $\chi^{(3)}$ nonlinearities, as a promising material for on-chip supercontinuum generation and frequency comb self-referencing. Broadband light from 500~nm to 4000 nm can be generated with only ${\sim}80$~mW (0.8~nJ) of 1560-nm pump power in the waveguide. Aluminum nitride provides an unexpected level of control over the output spectrum. In particular, the birefringence of the material enables a crossing of the TE and TM modes, which provides an enhancement in the spectral intensity by several orders of magnitude. In addition, we observe phase-mismatched difference frequency generation across the 3500 to 5500~nm region, which, if phase-matched, could provide a useful mid-infrared light source. Moreover, fully phase-matched second and third harmonic generation provide narrowband light that is tunable across the visible region.

Simultaneous second harmonic and supercontinuum generation processes allowed for the simplified detection of \fceo using a single, monolithic waveguide, and enabled high-quality stabilization of a compact laser frequency comb. In conclusion, aluminum nitride waveguides provide both robust comb stabilization as well as access to broad spectra across the visible, near infrared, and mid-infrared regions. These capabilities are crucial ingredients for building inexpensive, portable frequency combs for field applications, such as dual comb spectroscopy, spectrograph calibration, and precision metrology. 

\begin{acknowledgments}
The authors thank Nima Nader, Jeff Chiles, Frank Quinlan, and Tara Fortier for helpful discussions, and acknowledge assistance in device fabrication provided by Yale cleanroom staff Michael Power and Michael Rooks. 

This material is based upon work supported by the Air Force Office of Scientific Research under award number FA9550-16-1-0016, 
the Defense Advanced Research Projects Agency (DARPA) ACES, PULSE and SCOUT programs, 
the National Aeronautics and Space Administration (NASA), 
the National Institute of Standards and Technology (NIST), 
the National Research Council (NRC),
and the National Science Foundation (NSF) Graduate Research Fellowship Program (GRFP).

Certain commercial equipment, instruments, or materials are identified in this paper in order to specify the experimental procedure adequately. Such identification is not intended to imply recommendation or endorsement by the National Institute of Standards and Technology, nor is it intended to imply that the materials or equipment identified are necessarily the best available for the purpose. 

This work is a contribution of the United States government and is not subject to copyright in the United States of America.
\end{acknowledgments}

\bibliography{Zotero}

\begin{thebibliography}{56}%
\makeatletter
\providecommand \@ifxundefined [1]{%
 \@ifx{#1\undefined}
}%
\providecommand \@ifnum [1]{%
 \ifnum #1\expandafter \@firstoftwo
 \else \expandafter \@secondoftwo
 \fi
}%
\providecommand \@ifx [1]{%
 \ifx #1\expandafter \@firstoftwo
 \else \expandafter \@secondoftwo
 \fi
}%
\providecommand \natexlab [1]{#1}%
\providecommand \enquote  [1]{``#1''}%
\providecommand \bibnamefont  [1]{#1}%
\providecommand \bibfnamefont [1]{#1}%
\providecommand \citenamefont [1]{#1}%
\providecommand \href@noop [0]{\@secondoftwo}%
\providecommand \href [0]{\begingroup \@sanitize@url \@href}%
\providecommand \@href[1]{\@@startlink{#1}\@@href}%
\providecommand \@@href[1]{\endgroup#1\@@endlink}%
\providecommand \@sanitize@url [0]{\catcode `\\12\catcode `\$12\catcode
  `\&12\catcode `\#12\catcode `\^12\catcode `\_12\catcode `\%12\relax}%
\providecommand \@@startlink[1]{}%
\providecommand \@@endlink[0]{}%
\providecommand \url  [0]{\begingroup\@sanitize@url \@url }%
\providecommand \@url [1]{\endgroup\@href {#1}{\urlprefix }}%
\providecommand \urlprefix  [0]{URL }%
\providecommand \Eprint [0]{\href }%
\providecommand \doibase [0]{http://dx.doi.org/}%
\providecommand \selectlanguage [0]{\@gobble}%
\providecommand \bibinfo  [0]{\@secondoftwo}%
\providecommand \bibfield  [0]{\@secondoftwo}%
\providecommand \translation [1]{[#1]}%
\providecommand \BibitemOpen [0]{}%
\providecommand \bibitemStop [0]{}%
\providecommand \bibitemNoStop [0]{.\EOS\space}%
\providecommand \EOS [0]{\spacefactor3000\relax}%
\providecommand \BibitemShut  [1]{\csname bibitem#1\endcsname}%
\let\auto@bib@innerbib\@empty
\bibitem [{\citenamefont {Rosenband}\ \emph {et~al.}(2008)\citenamefont
  {Rosenband}, \citenamefont {Hume}, \citenamefont {Schmidt}, \citenamefont
  {Chou}, \citenamefont {Brusch}, \citenamefont {Lorini}, \citenamefont
  {Oskay}, \citenamefont {Drullinger}, \citenamefont {Fortier}, \citenamefont
  {Stalnaker}, \citenamefont {Diddams}, \citenamefont {Swann}, \citenamefont
  {Newbury}, \citenamefont {Itano}, \citenamefont {Wineland},\ and\
  \citenamefont {Bergquist}}]{rosenband_frequency_2008}%
  \BibitemOpen
  \bibfield  {author} {\bibinfo {author} {\bibfnamefont {T.}~\bibnamefont
  {Rosenband}}, \bibinfo {author} {\bibfnamefont {D.~B.}\ \bibnamefont {Hume}},
  \bibinfo {author} {\bibfnamefont {P.~O.}\ \bibnamefont {Schmidt}}, \bibinfo
  {author} {\bibfnamefont {C.~W.}\ \bibnamefont {Chou}}, \bibinfo {author}
  {\bibfnamefont {A.}~\bibnamefont {Brusch}}, \bibinfo {author} {\bibfnamefont
  {L.}~\bibnamefont {Lorini}}, \bibinfo {author} {\bibfnamefont {W.~H.}\
  \bibnamefont {Oskay}}, \bibinfo {author} {\bibfnamefont {R.~E.}\ \bibnamefont
  {Drullinger}}, \bibinfo {author} {\bibfnamefont {T.~M.}\ \bibnamefont
  {Fortier}}, \bibinfo {author} {\bibfnamefont {J.~E.}\ \bibnamefont
  {Stalnaker}}, \bibinfo {author} {\bibfnamefont {S.~A.}\ \bibnamefont
  {Diddams}}, \bibinfo {author} {\bibfnamefont {W.~C.}\ \bibnamefont {Swann}},
  \bibinfo {author} {\bibfnamefont {N.~R.}\ \bibnamefont {Newbury}}, \bibinfo
  {author} {\bibfnamefont {W.~M.}\ \bibnamefont {Itano}}, \bibinfo {author}
  {\bibfnamefont {D.~J.}\ \bibnamefont {Wineland}}, \ and\ \bibinfo {author}
  {\bibfnamefont {J.~C.}\ \bibnamefont {Bergquist}},\ }\bibfield  {title}
  {\enquote {\bibinfo {title} {Frequency {Ratio} of {Al}$^{\textrm{+}}$ and
  {Hg}$^{\textrm{+}}$ {Single}-{Ion} {Optical} {Clocks}; {Metrology} at the
  17th {Decimal} {Place}},}\ }\href {\doibase 10.1126/science.1154622}
  {\bibfield  {journal} {\bibinfo  {journal} {Science}\ }\textbf {\bibinfo
  {volume} {319}},\ \bibinfo {pages} {1808--1812} (\bibinfo {year}
  {2008})}\BibitemShut {NoStop}%
\bibitem [{\citenamefont {Rieker}\ \emph {et~al.}(2014)\citenamefont {Rieker},
  \citenamefont {Giorgetta}, \citenamefont {Swann}, \citenamefont {Kofler},
  \citenamefont {Zolot}, \citenamefont {Sinclair}, \citenamefont {Baumann},
  \citenamefont {Cromer}, \citenamefont {Petron}, \citenamefont {Sweeney},
  \citenamefont {Tans}, \citenamefont {Coddington},\ and\ \citenamefont
  {Newbury}}]{rieker_frequency-comb-based_2014}%
  \BibitemOpen
  \bibfield  {author} {\bibinfo {author} {\bibfnamefont {G.~B.}\ \bibnamefont
  {Rieker}}, \bibinfo {author} {\bibfnamefont {F.~R.}\ \bibnamefont
  {Giorgetta}}, \bibinfo {author} {\bibfnamefont {W.~C.}\ \bibnamefont
  {Swann}}, \bibinfo {author} {\bibfnamefont {J.}~\bibnamefont {Kofler}},
  \bibinfo {author} {\bibfnamefont {A.~M.}\ \bibnamefont {Zolot}}, \bibinfo
  {author} {\bibfnamefont {L.~C.}\ \bibnamefont {Sinclair}}, \bibinfo {author}
  {\bibfnamefont {E.}~\bibnamefont {Baumann}}, \bibinfo {author} {\bibfnamefont
  {C.}~\bibnamefont {Cromer}}, \bibinfo {author} {\bibfnamefont
  {G.}~\bibnamefont {Petron}}, \bibinfo {author} {\bibfnamefont
  {C.}~\bibnamefont {Sweeney}}, \bibinfo {author} {\bibfnamefont {P.~P.}\
  \bibnamefont {Tans}}, \bibinfo {author} {\bibfnamefont {I.}~\bibnamefont
  {Coddington}}, \ and\ \bibinfo {author} {\bibfnamefont {N.~R.}\ \bibnamefont
  {Newbury}},\ }\bibfield  {title} {\enquote {\bibinfo {title}
  {Frequency-comb-based remote sensing of greenhouse gases over kilometer air
  paths},}\ }\href {\doibase 10.1364/OPTICA.1.000290} {\bibfield  {journal}
  {\bibinfo  {journal} {Optica}\ }\textbf {\bibinfo {volume} {1}},\ \bibinfo
  {pages} {290--298} (\bibinfo {year} {2014})}\BibitemShut {NoStop}%
\bibitem [{\citenamefont {Waxman}\ \emph {et~al.}(2017)\citenamefont {Waxman},
  \citenamefont {Cossel}, \citenamefont {Truong}, \citenamefont {Giorgetta},
  \citenamefont {Swann}, \citenamefont {Coburn}, \citenamefont {Wright},
  \citenamefont {Rieker}, \citenamefont {Coddington},\ and\ \citenamefont
  {Newbury}}]{waxman_intercomparison_2017}%
  \BibitemOpen
  \bibfield  {author} {\bibinfo {author} {\bibfnamefont {E.~M.}\ \bibnamefont
  {Waxman}}, \bibinfo {author} {\bibfnamefont {K.~C.}\ \bibnamefont {Cossel}},
  \bibinfo {author} {\bibfnamefont {G.-W.}\ \bibnamefont {Truong}}, \bibinfo
  {author} {\bibfnamefont {F.~R.}\ \bibnamefont {Giorgetta}}, \bibinfo {author}
  {\bibfnamefont {W.~C.}\ \bibnamefont {Swann}}, \bibinfo {author}
  {\bibfnamefont {S.}~\bibnamefont {Coburn}}, \bibinfo {author} {\bibfnamefont
  {R.~J.}\ \bibnamefont {Wright}}, \bibinfo {author} {\bibfnamefont {G.~B.}\
  \bibnamefont {Rieker}}, \bibinfo {author} {\bibfnamefont {I.}~\bibnamefont
  {Coddington}}, \ and\ \bibinfo {author} {\bibfnamefont {N.~R.}\ \bibnamefont
  {Newbury}},\ }\bibfield  {title} {\enquote {\bibinfo {title} {Intercomparison
  of {Open}-{Path} {Trace} {Gas} {Measurements} with {Two} {Dual} {Frequency}
  {Comb} {Spectrometers}},}\ }\href {\doibase 10.5194/amt-2017-62} {\bibfield
  {journal} {\bibinfo  {journal} {Atmos. Meas. Tech. Discuss.}\ }\textbf
  {\bibinfo {volume} {2017}},\ \bibinfo {pages} {1--26} (\bibinfo {year}
  {2017})}\BibitemShut {NoStop}%
\bibitem [{\citenamefont {Li}\ \emph {et~al.}(2008)\citenamefont {Li},
  \citenamefont {Benedick}, \citenamefont {Fendel}, \citenamefont {Glenday},
  \citenamefont {Kärtner}, \citenamefont {Phillips}, \citenamefont {Sasselov},
  \citenamefont {Szentgyorgyi},\ and\ \citenamefont
  {Walsworth}}]{li_laser_2008}%
  \BibitemOpen
  \bibfield  {author} {\bibinfo {author} {\bibfnamefont {Chih-Hao}\
  \bibnamefont {Li}}, \bibinfo {author} {\bibfnamefont {Andrew~J.}\
  \bibnamefont {Benedick}}, \bibinfo {author} {\bibfnamefont {Peter}\
  \bibnamefont {Fendel}}, \bibinfo {author} {\bibfnamefont {Alexander~G.}\
  \bibnamefont {Glenday}}, \bibinfo {author} {\bibfnamefont {Franz~X.}\
  \bibnamefont {Kärtner}}, \bibinfo {author} {\bibfnamefont {David~F.}\
  \bibnamefont {Phillips}}, \bibinfo {author} {\bibfnamefont {Dimitar}\
  \bibnamefont {Sasselov}}, \bibinfo {author} {\bibfnamefont {Andrew}\
  \bibnamefont {Szentgyorgyi}}, \ and\ \bibinfo {author} {\bibfnamefont
  {Ronald~L.}\ \bibnamefont {Walsworth}},\ }\bibfield  {title} {\enquote
  {\bibinfo {title} {A laser frequency comb that enables radial velocity
  measurements with a precision of 1 cm s$^{\textrm{-1}}$},}\ }\href
  {\doibase 10.1038/nature06854} {\bibfield  {journal} {\bibinfo  {journal}
  {Nature}\ }\textbf {\bibinfo {volume} {452}},\ \bibinfo {pages} {610--612}
  (\bibinfo {year} {2008})}\BibitemShut {NoStop}%
\bibitem [{\citenamefont {Ycas}\ \emph {et~al.}(2012)\citenamefont {Ycas},
  \citenamefont {Quinlan}, \citenamefont {Diddams}, \citenamefont {Osterman},
  \citenamefont {Mahadevan}, \citenamefont {Redman}, \citenamefont {Terrien},
  \citenamefont {Ramsey}, \citenamefont {Bender}, \citenamefont {Botzer},\ and\
  \citenamefont {Sigurdsson}}]{ycas_demonstration_2012}%
  \BibitemOpen
  \bibfield  {author} {\bibinfo {author} {\bibfnamefont {Gabriel~G.}\
  \bibnamefont {Ycas}}, \bibinfo {author} {\bibfnamefont {Franklyn}\
  \bibnamefont {Quinlan}}, \bibinfo {author} {\bibfnamefont {Scott~A.}\
  \bibnamefont {Diddams}}, \bibinfo {author} {\bibfnamefont {Steve}\
  \bibnamefont {Osterman}}, \bibinfo {author} {\bibfnamefont {Suvrath}\
  \bibnamefont {Mahadevan}}, \bibinfo {author} {\bibfnamefont {Stephen}\
  \bibnamefont {Redman}}, \bibinfo {author} {\bibfnamefont {Ryan}\ \bibnamefont
  {Terrien}}, \bibinfo {author} {\bibfnamefont {Lawrence}\ \bibnamefont
  {Ramsey}}, \bibinfo {author} {\bibfnamefont {Chad~F.}\ \bibnamefont
  {Bender}}, \bibinfo {author} {\bibfnamefont {Brandon}\ \bibnamefont
  {Botzer}}, \ and\ \bibinfo {author} {\bibfnamefont {Steinn}\ \bibnamefont
  {Sigurdsson}},\ }\bibfield  {title} {\enquote {\bibinfo {title}
  {Demonstration of on-sky calibration of astronomical spectra using a 25 {GHz}
  near-{IR} laser frequency comb},}\ }\href {\doibase 10.1364/OE.20.006631}
  {\bibfield  {journal} {\bibinfo  {journal} {Optics Express}\ }\textbf
  {\bibinfo {volume} {20}},\ \bibinfo {pages} {6631--6643} (\bibinfo {year}
  {2012})}\BibitemShut {NoStop}%
\bibitem [{\citenamefont {Kippenberg}\ \emph {et~al.}(2011)\citenamefont
  {Kippenberg}, \citenamefont {Holzwarth},\ and\ \citenamefont
  {Diddams}}]{kippenberg_microresonator-based_2011}%
  \BibitemOpen
  \bibfield  {author} {\bibinfo {author} {\bibfnamefont {T.~J.}\ \bibnamefont
  {Kippenberg}}, \bibinfo {author} {\bibfnamefont {R.}~\bibnamefont
  {Holzwarth}}, \ and\ \bibinfo {author} {\bibfnamefont {S.~A.}\ \bibnamefont
  {Diddams}},\ }\bibfield  {title} {\enquote {\bibinfo {title}
  {Microresonator-{Based} {Optical} {Frequency} {Combs}},}\ }\href {\doibase
  10.1126/science.1193968} {\bibfield  {journal} {\bibinfo  {journal}
  {Science}\ }\textbf {\bibinfo {volume} {332}},\ \bibinfo {pages} {555--559}
  (\bibinfo {year} {2011})}\BibitemShut {NoStop}%
\bibitem [{\citenamefont {Dudley}\ \emph {et~al.}(2006)\citenamefont {Dudley},
  \citenamefont {Genty},\ and\ \citenamefont
  {Coen}}]{dudley_supercontinuum_2006}%
  \BibitemOpen
  \bibfield  {author} {\bibinfo {author} {\bibfnamefont {John~M.}\ \bibnamefont
  {Dudley}}, \bibinfo {author} {\bibfnamefont {Goëry}\ \bibnamefont {Genty}},
  \ and\ \bibinfo {author} {\bibfnamefont {Stéphane}\ \bibnamefont {Coen}},\
  }\bibfield  {title} {\enquote {\bibinfo {title} {Supercontinuum generation in
  photonic crystal fiber},}\ }\href {\doibase 10.1103/RevModPhys.78.1135}
  {\bibfield  {journal} {\bibinfo  {journal} {Reviews of Modern Physics}\
  }\textbf {\bibinfo {volume} {78}},\ \bibinfo {pages} {1135--1184} (\bibinfo
  {year} {2006})}\BibitemShut {NoStop}%
\bibitem [{\citenamefont {Jones}\ \emph {et~al.}(2000)\citenamefont {Jones},
  \citenamefont {Diddams}, \citenamefont {Ranka}, \citenamefont {Stentz},
  \citenamefont {Windeler}, \citenamefont {Hall},\ and\ \citenamefont
  {Cundiff}}]{jones_carrier-envelope_2000}%
  \BibitemOpen
  \bibfield  {author} {\bibinfo {author} {\bibfnamefont {David~J.}\
  \bibnamefont {Jones}}, \bibinfo {author} {\bibfnamefont {Scott~A.}\
  \bibnamefont {Diddams}}, \bibinfo {author} {\bibfnamefont {Jinendra~K.}\
  \bibnamefont {Ranka}}, \bibinfo {author} {\bibfnamefont {Andrew}\
  \bibnamefont {Stentz}}, \bibinfo {author} {\bibfnamefont {Robert~S.}\
  \bibnamefont {Windeler}}, \bibinfo {author} {\bibfnamefont {John~L.}\
  \bibnamefont {Hall}}, \ and\ \bibinfo {author} {\bibfnamefont {Steven~T.}\
  \bibnamefont {Cundiff}},\ }\bibfield  {title} {\enquote {\bibinfo {title}
  {Carrier-{Envelope} {Phase} {Control} of {Femtosecond} {Mode}-{Locked}
  {Lasers} and {Direct} {Optical} {Frequency} {Synthesis}},}\ }\href {\doibase
  10.1126/science.288.5466.635} {\bibfield  {journal} {\bibinfo  {journal}
  {Science}\ }\textbf {\bibinfo {volume} {288}},\ \bibinfo {pages} {635--639}
  (\bibinfo {year} {2000})}\BibitemShut {NoStop}%
\bibitem [{\citenamefont {Holzwarth}\ \emph {et~al.}(2000)\citenamefont
  {Holzwarth}, \citenamefont {Udem}, \citenamefont {Hänsch}, \citenamefont
  {Knight}, \citenamefont {Wadsworth},\ and\ \citenamefont
  {Russell}}]{holzwarth_optical_2000}%
  \BibitemOpen
  \bibfield  {author} {\bibinfo {author} {\bibfnamefont {R.}~\bibnamefont
  {Holzwarth}}, \bibinfo {author} {\bibfnamefont {Th.}\ \bibnamefont {Udem}},
  \bibinfo {author} {\bibfnamefont {T.~W.}\ \bibnamefont {Hänsch}}, \bibinfo
  {author} {\bibfnamefont {J.~C.}\ \bibnamefont {Knight}}, \bibinfo {author}
  {\bibfnamefont {W.~J.}\ \bibnamefont {Wadsworth}}, \ and\ \bibinfo {author}
  {\bibfnamefont {P.~St.~J.}\ \bibnamefont {Russell}},\ }\bibfield  {title}
  {\enquote {\bibinfo {title} {Optical {Frequency} {Synthesizer} for
  {Precision} {Spectroscopy}},}\ }\href {\doibase 10.1103/PhysRevLett.85.2264}
  {\bibfield  {journal} {\bibinfo  {journal} {Physical Review Letters}\
  }\textbf {\bibinfo {volume} {85}},\ \bibinfo {pages} {2264--2267} (\bibinfo
  {year} {2000})}\BibitemShut {NoStop}%
\bibitem [{\citenamefont {Diddams}\ \emph {et~al.}(2000)\citenamefont
  {Diddams}, \citenamefont {Jones}, \citenamefont {Ye}, \citenamefont
  {Cundiff}, \citenamefont {Hall}, \citenamefont {Ranka}, \citenamefont
  {Windeler}, \citenamefont {Holzwarth}, \citenamefont {Udem},\ and\
  \citenamefont {Hänsch}}]{diddams_direct_2000}%
  \BibitemOpen
  \bibfield  {author} {\bibinfo {author} {\bibfnamefont {Scott~A.}\
  \bibnamefont {Diddams}}, \bibinfo {author} {\bibfnamefont {David~J.}\
  \bibnamefont {Jones}}, \bibinfo {author} {\bibfnamefont {Jun}\ \bibnamefont
  {Ye}}, \bibinfo {author} {\bibfnamefont {Steven~T.}\ \bibnamefont {Cundiff}},
  \bibinfo {author} {\bibfnamefont {John~L.}\ \bibnamefont {Hall}}, \bibinfo
  {author} {\bibfnamefont {Jinendra~K.}\ \bibnamefont {Ranka}}, \bibinfo
  {author} {\bibfnamefont {Robert~S.}\ \bibnamefont {Windeler}}, \bibinfo
  {author} {\bibfnamefont {Ronald}\ \bibnamefont {Holzwarth}}, \bibinfo
  {author} {\bibfnamefont {Thomas}\ \bibnamefont {Udem}}, \ and\ \bibinfo
  {author} {\bibfnamefont {T.~W.}\ \bibnamefont {Hänsch}},\ }\bibfield
  {title} {\enquote {\bibinfo {title} {Direct {Link} between {Microwave} and
  {Optical} {Frequencies} with a 300 {THz} {Femtosecond} {Laser} {Comb}},}\
  }\href {\doibase 10.1103/PhysRevLett.84.5102} {\bibfield  {journal} {\bibinfo
   {journal} {Physical Review Letters}\ }\textbf {\bibinfo {volume} {84}},\
  \bibinfo {pages} {5102--5105} (\bibinfo {year} {2000})}\BibitemShut {NoStop}%
\bibitem [{\citenamefont {Kobayashi}\ \emph {et~al.}(1972)\citenamefont
  {Kobayashi}, \citenamefont {Sueta}, \citenamefont {Cho},\ and\ \citenamefont
  {Matsu}}]{kobayashi_highrepetitionrate_1972}%
  \BibitemOpen
  \bibfield  {author} {\bibinfo {author} {\bibfnamefont {T.}~\bibnamefont
  {Kobayashi}}, \bibinfo {author} {\bibfnamefont {T}~\bibnamefont {Sueta}},
  \bibinfo {author} {\bibfnamefont {Y.}~\bibnamefont {Cho}}, \ and\ \bibinfo
  {author} {\bibfnamefont {Y.}~\bibnamefont {Matsu}},\ }\bibfield  {title}
  {\enquote {\bibinfo {title} {High‐repetition‐rate optical pulse generator
  using a {Fabry}‐{Perot} electro‐optic modulator},}\ }\href {\doibase
  10.1063/1.1654403} {\bibfield  {journal} {\bibinfo  {journal} {Applied
  Physics Letters}\ }\textbf {\bibinfo {volume} {21}},\ \bibinfo {pages}
  {341--343} (\bibinfo {year} {1972})}\BibitemShut {NoStop}%
\bibitem [{\citenamefont {Torres-Company}\ and\ \citenamefont
  {Weiner}(2014)}]{torres-company_optical_2014}%
  \BibitemOpen
  \bibfield  {author} {\bibinfo {author} {\bibfnamefont {Victor}\ \bibnamefont
  {Torres-Company}}\ and\ \bibinfo {author} {\bibfnamefont {Andrew~M.}\
  \bibnamefont {Weiner}},\ }\bibfield  {title} {\enquote {\bibinfo {title}
  {Optical frequency comb technology for ultra-broadband radio-frequency
  photonics},}\ }\href {\doibase 10.1002/lpor.201300126} {\bibfield  {journal}
  {\bibinfo  {journal} {Laser \& Photonics Reviews}\ }\textbf {\bibinfo
  {volume} {8}},\ \bibinfo {pages} {368--393} (\bibinfo {year}
  {2014})}\BibitemShut {NoStop}%
\bibitem [{\citenamefont {Del’Haye}\ \emph {et~al.}(2007)\citenamefont
  {Del’Haye}, \citenamefont {Schliesser}, \citenamefont {Arcizet},
  \citenamefont {Wilken}, \citenamefont {Holzwarth},\ and\ \citenamefont
  {Kippenberg}}]{delhaye_optical_2007}%
  \BibitemOpen
  \bibfield  {author} {\bibinfo {author} {\bibfnamefont {P.}~\bibnamefont
  {Del’Haye}}, \bibinfo {author} {\bibfnamefont {A.}~\bibnamefont
  {Schliesser}}, \bibinfo {author} {\bibfnamefont {O.}~\bibnamefont {Arcizet}},
  \bibinfo {author} {\bibfnamefont {T.}~\bibnamefont {Wilken}}, \bibinfo
  {author} {\bibfnamefont {R.}~\bibnamefont {Holzwarth}}, \ and\ \bibinfo
  {author} {\bibfnamefont {T.~J.}\ \bibnamefont {Kippenberg}},\ }\bibfield
  {title} {\enquote {\bibinfo {title} {Optical frequency comb generation from a
  monolithic microresonator},}\ }\href {\doibase 10.1038/nature06401}
  {\bibfield  {journal} {\bibinfo  {journal} {Nature}\ }\textbf {\bibinfo
  {volume} {450}},\ \bibinfo {pages} {1214--1217} (\bibinfo {year}
  {2007})}\BibitemShut {NoStop}%
\bibitem [{\citenamefont {Herr}\ \emph
  {et~al.}(2014{\natexlab{a}})\citenamefont {Herr}, \citenamefont {Brasch},
  \citenamefont {Jost}, \citenamefont {Wang}, \citenamefont {Kondratiev},
  \citenamefont {Gorodetsky},\ and\ \citenamefont
  {Kippenberg}}]{herr_temporal_2014}%
  \BibitemOpen
  \bibfield  {author} {\bibinfo {author} {\bibfnamefont {T.}~\bibnamefont
  {Herr}}, \bibinfo {author} {\bibfnamefont {V.}~\bibnamefont {Brasch}},
  \bibinfo {author} {\bibfnamefont {J.~D.}\ \bibnamefont {Jost}}, \bibinfo
  {author} {\bibfnamefont {C.~Y.}\ \bibnamefont {Wang}}, \bibinfo {author}
  {\bibfnamefont {N.~M.}\ \bibnamefont {Kondratiev}}, \bibinfo {author}
  {\bibfnamefont {M.~L.}\ \bibnamefont {Gorodetsky}}, \ and\ \bibinfo {author}
  {\bibfnamefont {T.~J.}\ \bibnamefont {Kippenberg}},\ }\bibfield  {title}
  {\enquote {\bibinfo {title} {Temporal solitons in optical microresonators},}\
  }\href {\doibase 10.1038/nphoton.2013.343} {\bibfield  {journal} {\bibinfo
  {journal} {Nature Photonics}\ }\textbf {\bibinfo {volume} {8}},\ \bibinfo
  {pages} {145--152} (\bibinfo {year} {2014}{\natexlab{a}})}\BibitemShut
  {NoStop}%
\bibitem [{\citenamefont {Schliesser}\ \emph {et~al.}(2012)\citenamefont
  {Schliesser}, \citenamefont {Picque},\ and\ \citenamefont
  {Hansch}}]{schliesser_mid-infrared_2012}%
  \BibitemOpen
  \bibfield  {author} {\bibinfo {author} {\bibfnamefont {Albert}\ \bibnamefont
  {Schliesser}}, \bibinfo {author} {\bibfnamefont {Nathalie}\ \bibnamefont
  {Picque}}, \ and\ \bibinfo {author} {\bibfnamefont {Theodor~W.}\ \bibnamefont
  {Hansch}},\ }\bibfield  {title} {\enquote {\bibinfo {title} {Mid-infrared
  frequency combs},}\ }\href {\doibase 10.1038/nphoton.2012.142} {\bibfield
  {journal} {\bibinfo  {journal} {Nature Photonics}\ }\textbf {\bibinfo
  {volume} {6}},\ \bibinfo {pages} {440--449} (\bibinfo {year}
  {2012})}\BibitemShut {NoStop}%
\bibitem [{\citenamefont {Coddington}\ \emph {et~al.}(2016)\citenamefont
  {Coddington}, \citenamefont {Newbury},\ and\ \citenamefont
  {Swann}}]{coddington_dual-comb_2016}%
  \BibitemOpen
  \bibfield  {author} {\bibinfo {author} {\bibfnamefont {Ian}\ \bibnamefont
  {Coddington}}, \bibinfo {author} {\bibfnamefont {Nathan}\ \bibnamefont
  {Newbury}}, \ and\ \bibinfo {author} {\bibfnamefont {William}\ \bibnamefont
  {Swann}},\ }\bibfield  {title} {\enquote {\bibinfo {title} {Dual-comb
  spectroscopy},}\ }\href {\doibase 10.1364/OPTICA.3.000414} {\bibfield
  {journal} {\bibinfo  {journal} {Optica}\ }\textbf {\bibinfo {volume} {3}},\
  \bibinfo {pages} {414--426} (\bibinfo {year} {2016})}\BibitemShut {NoStop}%
\bibitem [{\citenamefont {Truong}\ \emph {et~al.}(2016)\citenamefont {Truong},
  \citenamefont {Waxman}, \citenamefont {Cossel}, \citenamefont {Giorgetta},
  \citenamefont {Swann}, \citenamefont {Coddington},\ and\ \citenamefont
  {Newbury}}]{truong_dual-comb_2016}%
  \BibitemOpen
  \bibfield  {author} {\bibinfo {author} {\bibfnamefont {Gar-Wing}\
  \bibnamefont {Truong}}, \bibinfo {author} {\bibfnamefont {Eleanor}\
  \bibnamefont {Waxman}}, \bibinfo {author} {\bibfnamefont {Kevin~C.}\
  \bibnamefont {Cossel}}, \bibinfo {author} {\bibfnamefont {Fabrizio}\
  \bibnamefont {Giorgetta}}, \bibinfo {author} {\bibfnamefont {William~C.}\
  \bibnamefont {Swann}}, \bibinfo {author} {\bibfnamefont {Ian~R.}\
  \bibnamefont {Coddington}}, \ and\ \bibinfo {author} {\bibfnamefont
  {Nathan~R.}\ \bibnamefont {Newbury}},\ }\bibfield  {title} {\enquote
  {\bibinfo {title} {Dual-comb {Spectroscopy} for {City}-scale {Open} {Path}
  {Greenhouse} {Gas} {Monitoring}},}\ }in\ \href {\doibase
  10.1364/CLEO_SI.2016.SW4H.2} {\emph {\bibinfo {booktitle} {Conference on
  {Lasers} and {Electro}-{Optics} (2016), paper {SW}4H.2}}}\ (\bibinfo
  {publisher} {Optical Society of America},\ \bibinfo {year} {2016})\ p.\
  \bibinfo {pages} {SW4H.2}\BibitemShut {NoStop}%
\bibitem [{\citenamefont {Giorgetta}\ \emph {et~al.}(2015)\citenamefont
  {Giorgetta}, \citenamefont {Rieker}, \citenamefont {Baumann}, \citenamefont
  {Swann}, \citenamefont {Sinclair}, \citenamefont {Kofler}, \citenamefont
  {Coddington},\ and\ \citenamefont {Newbury}}]{giorgetta_broadband_2015}%
  \BibitemOpen
  \bibfield  {author} {\bibinfo {author} {\bibfnamefont {Fabrizio~R.}\
  \bibnamefont {Giorgetta}}, \bibinfo {author} {\bibfnamefont {Gregory~B.}\
  \bibnamefont {Rieker}}, \bibinfo {author} {\bibfnamefont {Esther}\
  \bibnamefont {Baumann}}, \bibinfo {author} {\bibfnamefont {William~C.}\
  \bibnamefont {Swann}}, \bibinfo {author} {\bibfnamefont {Laura~C.}\
  \bibnamefont {Sinclair}}, \bibinfo {author} {\bibfnamefont {Jon}\
  \bibnamefont {Kofler}}, \bibinfo {author} {\bibfnamefont {Ian}\ \bibnamefont
  {Coddington}}, \ and\ \bibinfo {author} {\bibfnamefont {Nathan~R.}\
  \bibnamefont {Newbury}},\ }\bibfield  {title} {\enquote {\bibinfo {title}
  {Broadband {Phase} {Spectroscopy} over {Turbulent} {Air} {Paths}},}\ }\href
  {\doibase 10.1103/PhysRevLett.115.103901} {\bibfield  {journal} {\bibinfo
  {journal} {Physical Review Letters}\ }\textbf {\bibinfo {volume} {115}},\
  \bibinfo {pages} {103901} (\bibinfo {year} {2015})}\BibitemShut {NoStop}%
\bibitem [{\citenamefont {Cossel}\ \emph {et~al.}(2017)\citenamefont {Cossel},
  \citenamefont {Waxman}, \citenamefont {Finneran}, \citenamefont {Blake},
  \citenamefont {Ye},\ and\ \citenamefont {Newbury}}]{cossel_gas-phase_2017}%
  \BibitemOpen
  \bibfield  {author} {\bibinfo {author} {\bibfnamefont {Kevin~C.}\
  \bibnamefont {Cossel}}, \bibinfo {author} {\bibfnamefont {Eleanor~M.}\
  \bibnamefont {Waxman}}, \bibinfo {author} {\bibfnamefont {Ian~A.}\
  \bibnamefont {Finneran}}, \bibinfo {author} {\bibfnamefont {Geoffrey~A.}\
  \bibnamefont {Blake}}, \bibinfo {author} {\bibfnamefont {Jun}\ \bibnamefont
  {Ye}}, \ and\ \bibinfo {author} {\bibfnamefont {Nathan~R.}\ \bibnamefont
  {Newbury}},\ }\bibfield  {title} {\enquote {\bibinfo {title} {Gas-phase
  broadband spectroscopy using active sources: progress, status, and
  applications [{Invited}]},}\ }\href {\doibase 10.1364/JOSAB.34.000104}
  {\bibfield  {journal} {\bibinfo  {journal} {JOSA B}\ }\textbf {\bibinfo
  {volume} {34}},\ \bibinfo {pages} {104--129} (\bibinfo {year}
  {2017})}\BibitemShut {NoStop}%
\bibitem [{\citenamefont {Epping}\ \emph {et~al.}(2015)\citenamefont {Epping},
  \citenamefont {Hellwig}, \citenamefont {Hoekman}, \citenamefont {Mateman},
  \citenamefont {Leinse}, \citenamefont {Heideman}, \citenamefont {Rees},
  \citenamefont {Slot}, \citenamefont {Lee}, \citenamefont {Fallnich},\ and\
  \citenamefont {Boller}}]{epping_chip_2015}%
  \BibitemOpen
  \bibfield  {author} {\bibinfo {author} {\bibfnamefont {Jorn~P.}\ \bibnamefont
  {Epping}}, \bibinfo {author} {\bibfnamefont {Tim}\ \bibnamefont {Hellwig}},
  \bibinfo {author} {\bibfnamefont {Marcel}\ \bibnamefont {Hoekman}}, \bibinfo
  {author} {\bibfnamefont {Richard}\ \bibnamefont {Mateman}}, \bibinfo {author}
  {\bibfnamefont {Arne}\ \bibnamefont {Leinse}}, \bibinfo {author}
  {\bibfnamefont {René~G.}\ \bibnamefont {Heideman}}, \bibinfo {author}
  {\bibfnamefont {Albert~van}\ \bibnamefont {Rees}}, \bibinfo {author}
  {\bibfnamefont {Peter J. M. van~der}\ \bibnamefont {Slot}}, \bibinfo {author}
  {\bibfnamefont {Chris~J.}\ \bibnamefont {Lee}}, \bibinfo {author}
  {\bibfnamefont {Carsten}\ \bibnamefont {Fallnich}}, \ and\ \bibinfo {author}
  {\bibfnamefont {Klaus-J.}\ \bibnamefont {Boller}},\ }\bibfield  {title}
  {\enquote {\bibinfo {title} {On chip visible-to-infrared supercontinuum
  generation with more than 495 {THz} spectral bandwidth},}\ }\href {\doibase
  10.1364/OE.23.019596} {\bibfield  {journal} {\bibinfo  {journal} {Optics
  Express}\ }\textbf {\bibinfo {volume} {23}},\ \bibinfo {pages} {19596--19604}
  (\bibinfo {year} {2015})}\BibitemShut {NoStop}%
\bibitem [{\citenamefont {Porcel}\ \emph {et~al.}(2017)\citenamefont {Porcel},
  \citenamefont {Schepers}, \citenamefont {Epping}, \citenamefont {Hellwig},
  \citenamefont {Hoekman}, \citenamefont {Heideman}, \citenamefont {Slot},
  \citenamefont {Lee}, \citenamefont {Schmidt}, \citenamefont {Bratschitsch},
  \citenamefont {Fallnich},\ and\ \citenamefont
  {Boller}}]{porcel_two-octave_2017}%
  \BibitemOpen
  \bibfield  {author} {\bibinfo {author} {\bibfnamefont {Marco A.~G.}\
  \bibnamefont {Porcel}}, \bibinfo {author} {\bibfnamefont {Florian}\
  \bibnamefont {Schepers}}, \bibinfo {author} {\bibfnamefont {Jörn~P.}\
  \bibnamefont {Epping}}, \bibinfo {author} {\bibfnamefont {Tim}\ \bibnamefont
  {Hellwig}}, \bibinfo {author} {\bibfnamefont {Marcel}\ \bibnamefont
  {Hoekman}}, \bibinfo {author} {\bibfnamefont {René~G.}\ \bibnamefont
  {Heideman}}, \bibinfo {author} {\bibfnamefont {Peter J. M. van~der}\
  \bibnamefont {Slot}}, \bibinfo {author} {\bibfnamefont {Chris~J.}\
  \bibnamefont {Lee}}, \bibinfo {author} {\bibfnamefont {Robert}\ \bibnamefont
  {Schmidt}}, \bibinfo {author} {\bibfnamefont {Rudolf}\ \bibnamefont
  {Bratschitsch}}, \bibinfo {author} {\bibfnamefont {Carsten}\ \bibnamefont
  {Fallnich}}, \ and\ \bibinfo {author} {\bibfnamefont {Klaus-J.}\ \bibnamefont
  {Boller}},\ }\bibfield  {title} {\enquote {\bibinfo {title} {Two-octave
  spanning supercontinuum generation in stoichiometric silicon nitride
  waveguides pumped at telecom wavelengths},}\ }\href {\doibase
  10.1364/OE.25.001542} {\bibfield  {journal} {\bibinfo  {journal} {Optics
  Express}\ }\textbf {\bibinfo {volume} {25}},\ \bibinfo {pages} {1542--1554}
  (\bibinfo {year} {2017})}\BibitemShut {NoStop}%
\bibitem [{\citenamefont {Klenner}\ \emph {et~al.}(2016)\citenamefont
  {Klenner}, \citenamefont {Mayer}, \citenamefont {Johnson}, \citenamefont
  {Luke}, \citenamefont {Lamont}, \citenamefont {Okawachi}, \citenamefont
  {Lipson}, \citenamefont {Gaeta},\ and\ \citenamefont
  {Keller}}]{klenner_gigahertz_2016}%
  \BibitemOpen
  \bibfield  {author} {\bibinfo {author} {\bibfnamefont {Alexander}\
  \bibnamefont {Klenner}}, \bibinfo {author} {\bibfnamefont {Aline~S.}\
  \bibnamefont {Mayer}}, \bibinfo {author} {\bibfnamefont {Adrea~R.}\
  \bibnamefont {Johnson}}, \bibinfo {author} {\bibfnamefont {Kevin}\
  \bibnamefont {Luke}}, \bibinfo {author} {\bibfnamefont {Michael R.~E.}\
  \bibnamefont {Lamont}}, \bibinfo {author} {\bibfnamefont {Yoshitomo}\
  \bibnamefont {Okawachi}}, \bibinfo {author} {\bibfnamefont {Michal}\
  \bibnamefont {Lipson}}, \bibinfo {author} {\bibfnamefont {Alexander~L.}\
  \bibnamefont {Gaeta}}, \ and\ \bibinfo {author} {\bibfnamefont {Ursula}\
  \bibnamefont {Keller}},\ }\bibfield  {title} {\enquote {\bibinfo {title}
  {Gigahertz frequency comb offset stabilization based on supercontinuum
  generation in silicon nitride waveguides},}\ }\href {\doibase
  10.1364/OE.24.011043} {\bibfield  {journal} {\bibinfo  {journal} {Optics
  Express}\ }\textbf {\bibinfo {volume} {24}},\ \bibinfo {pages} {11043--11053}
  (\bibinfo {year} {2016})}\BibitemShut {NoStop}%
\bibitem [{\citenamefont {Mayer}\ \emph {et~al.}(2015)\citenamefont {Mayer},
  \citenamefont {Klenner}, \citenamefont {Johnson}, \citenamefont {Luke},
  \citenamefont {Lamont}, \citenamefont {Okawachi}, \citenamefont {Lipson},
  \citenamefont {Gaeta},\ and\ \citenamefont {Keller}}]{mayer_frequency_2015}%
  \BibitemOpen
  \bibfield  {author} {\bibinfo {author} {\bibfnamefont {A.~S.}\ \bibnamefont
  {Mayer}}, \bibinfo {author} {\bibfnamefont {A.}~\bibnamefont {Klenner}},
  \bibinfo {author} {\bibfnamefont {A.~R.}\ \bibnamefont {Johnson}}, \bibinfo
  {author} {\bibfnamefont {K.}~\bibnamefont {Luke}}, \bibinfo {author}
  {\bibfnamefont {M.~R.~E.}\ \bibnamefont {Lamont}}, \bibinfo {author}
  {\bibfnamefont {Y.}~\bibnamefont {Okawachi}}, \bibinfo {author}
  {\bibfnamefont {M.}~\bibnamefont {Lipson}}, \bibinfo {author} {\bibfnamefont
  {A.~L.}\ \bibnamefont {Gaeta}}, \ and\ \bibinfo {author} {\bibfnamefont
  {U.}~\bibnamefont {Keller}},\ }\bibfield  {title} {\enquote {\bibinfo {title}
  {Frequency comb offset detection using supercontinuum generation in silicon
  nitride waveguides},}\ }\href {\doibase 10.1364/OE.23.015440} {\bibfield
  {journal} {\bibinfo  {journal} {Optics Express}\ }\textbf {\bibinfo {volume}
  {23}},\ \bibinfo {pages} {15440--15451} (\bibinfo {year} {2015})}\BibitemShut
  {NoStop}%
\bibitem [{\citenamefont {Boggio}\ \emph {et~al.}(2014)\citenamefont {Boggio},
  \citenamefont {Bodenmüller}, \citenamefont {Fremberg}, \citenamefont
  {Haynes}, \citenamefont {Roth}, \citenamefont {Eisermann}, \citenamefont
  {Lisker}, \citenamefont {Zimmermann},\ and\ \citenamefont
  {Böhm}}]{boggio_dispersion_2014}%
  \BibitemOpen
  \bibfield  {author} {\bibinfo {author} {\bibfnamefont {J.~M.~Chavez}\
  \bibnamefont {Boggio}}, \bibinfo {author} {\bibfnamefont {D.}~\bibnamefont
  {Bodenmüller}}, \bibinfo {author} {\bibfnamefont {T.}~\bibnamefont
  {Fremberg}}, \bibinfo {author} {\bibfnamefont {R.}~\bibnamefont {Haynes}},
  \bibinfo {author} {\bibfnamefont {M.~M.}\ \bibnamefont {Roth}}, \bibinfo
  {author} {\bibfnamefont {R.}~\bibnamefont {Eisermann}}, \bibinfo {author}
  {\bibfnamefont {M.}~\bibnamefont {Lisker}}, \bibinfo {author} {\bibfnamefont
  {L.}~\bibnamefont {Zimmermann}}, \ and\ \bibinfo {author} {\bibfnamefont
  {M.}~\bibnamefont {Böhm}},\ }\bibfield  {title} {\enquote {\bibinfo {title}
  {Dispersion engineered silicon nitride waveguides by geometrical and
  refractive-index optimization},}\ }\href {\doibase 10.1364/JOSAB.31.002846}
  {\bibfield  {journal} {\bibinfo  {journal} {JOSA B}\ }\textbf {\bibinfo
  {volume} {31}},\ \bibinfo {pages} {2846--2857} (\bibinfo {year}
  {2014})}\BibitemShut {NoStop}%
\bibitem [{\citenamefont {Hickstein}\ \emph {et~al.}(2016)\citenamefont
  {Hickstein}, \citenamefont {Ycas}, \citenamefont {Lind}, \citenamefont
  {Cole}, \citenamefont {Srinivasan}, \citenamefont {Diddams},\ and\
  \citenamefont {Papp}}]{hickstein_photonic-chip_2016}%
  \BibitemOpen
  \bibfield  {author} {\bibinfo {author} {\bibfnamefont {Daniel}\ \bibnamefont
  {Hickstein}}, \bibinfo {author} {\bibfnamefont {Gabriel}\ \bibnamefont
  {Ycas}}, \bibinfo {author} {\bibfnamefont {Alex}\ \bibnamefont {Lind}},
  \bibinfo {author} {\bibfnamefont {Daniel~C.}\ \bibnamefont {Cole}}, \bibinfo
  {author} {\bibfnamefont {Katrik}\ \bibnamefont {Srinivasan}}, \bibinfo
  {author} {\bibfnamefont {Scott}\ \bibnamefont {Diddams}}, \ and\ \bibinfo
  {author} {\bibfnamefont {Scott}\ \bibnamefont {Papp}},\ }\bibfield  {title}
  {\enquote {\bibinfo {title} {Photonic-chip {Waveguides} for {Supercontinuum}
  {Generation} with {Picojoule} {Pulses}},}\ }in\ \href {\doibase
  10.1364/IPRSN.2016.IM3A.2} {\emph {\bibinfo {booktitle} {Advanced {Photonics}
  2016 ({IPR}, {NOMA}, {Sensors}, {Networks}, {SPPCom}, {SOF}) (2016), paper
  {IM}3A.2}}}\ (\bibinfo  {publisher} {Optical Society of America},\ \bibinfo
  {year} {2016})\ p.\ \bibinfo {pages} {IM3A.2}\BibitemShut {NoStop}%
\bibitem [{\citenamefont {Carlson}\ \emph
  {et~al.}(2017{\natexlab{a}})\citenamefont {Carlson}, \citenamefont
  {Hickstein}, \citenamefont {Lind}, \citenamefont {Olson}, \citenamefont
  {Fox}, \citenamefont {Brown}, \citenamefont {Ludlow}, \citenamefont {Li},
  \citenamefont {Westly}, \citenamefont {Leopardi}, \citenamefont {Fortier},
  \citenamefont {Srinivasan}, \citenamefont {Diddams},\ and\ \citenamefont
  {Papp}}]{carlson_photonic-chip_2017}%
  \BibitemOpen
  \bibfield  {author} {\bibinfo {author} {\bibfnamefont {David}\ \bibnamefont
  {Carlson}}, \bibinfo {author} {\bibfnamefont {Daniel}\ \bibnamefont
  {Hickstein}}, \bibinfo {author} {\bibfnamefont {Alexander}\ \bibnamefont
  {Lind}}, \bibinfo {author} {\bibfnamefont {Judith}\ \bibnamefont {Olson}},
  \bibinfo {author} {\bibfnamefont {Richard}\ \bibnamefont {Fox}}, \bibinfo
  {author} {\bibfnamefont {Roger}\ \bibnamefont {Brown}}, \bibinfo {author}
  {\bibfnamefont {Andrew}\ \bibnamefont {Ludlow}}, \bibinfo {author}
  {\bibfnamefont {Qing}\ \bibnamefont {Li}}, \bibinfo {author} {\bibfnamefont
  {Daron}\ \bibnamefont {Westly}}, \bibinfo {author} {\bibfnamefont {Holly}\
  \bibnamefont {Leopardi}}, \bibinfo {author} {\bibfnamefont {Tara}\
  \bibnamefont {Fortier}}, \bibinfo {author} {\bibfnamefont {Kartik}\
  \bibnamefont {Srinivasan}}, \bibinfo {author} {\bibfnamefont {Scott}\
  \bibnamefont {Diddams}}, \ and\ \bibinfo {author} {\bibfnamefont {Scott}\
  \bibnamefont {Papp}},\ }\bibfield  {title} {\enquote {\bibinfo {title}
  {Photonic-chip supercontinuum with tailored spectra for precision frequency
  metrology},}\ }\href {http://arxiv.org/abs/1702.03269} {\bibfield  {journal}
  {\bibinfo  {journal} {arXiv:1702.03269 [physics]}\ } (\bibinfo {year}
  {2017}{\natexlab{a}})}\BibitemShut {NoStop}%
\bibitem [{\citenamefont {Johnson}\ \emph {et~al.}(2015)\citenamefont
  {Johnson}, \citenamefont {Mayer}, \citenamefont {Klenner}, \citenamefont
  {Luke}, \citenamefont {Lamb}, \citenamefont {Lamont}, \citenamefont {Joshi},
  \citenamefont {Okawachi}, \citenamefont {Wise}, \citenamefont {Lipson},
  \citenamefont {Keller},\ and\ \citenamefont
  {Gaeta}}]{johnson_octave-spanning_2015}%
  \BibitemOpen
  \bibfield  {author} {\bibinfo {author} {\bibfnamefont {Adrea~R.}\
  \bibnamefont {Johnson}}, \bibinfo {author} {\bibfnamefont {Aline~S.}\
  \bibnamefont {Mayer}}, \bibinfo {author} {\bibfnamefont {Alexander}\
  \bibnamefont {Klenner}}, \bibinfo {author} {\bibfnamefont {Kevin}\
  \bibnamefont {Luke}}, \bibinfo {author} {\bibfnamefont {Erin~S.}\
  \bibnamefont {Lamb}}, \bibinfo {author} {\bibfnamefont {Michael R.~E.}\
  \bibnamefont {Lamont}}, \bibinfo {author} {\bibfnamefont {Chaitanya}\
  \bibnamefont {Joshi}}, \bibinfo {author} {\bibfnamefont {Yoshitomo}\
  \bibnamefont {Okawachi}}, \bibinfo {author} {\bibfnamefont {Frank~W.}\
  \bibnamefont {Wise}}, \bibinfo {author} {\bibfnamefont {Michal}\ \bibnamefont
  {Lipson}}, \bibinfo {author} {\bibfnamefont {Ursula}\ \bibnamefont {Keller}},
  \ and\ \bibinfo {author} {\bibfnamefont {Alexander~L.}\ \bibnamefont
  {Gaeta}},\ }\bibfield  {title} {\enquote {\bibinfo {title} {Octave-spanning
  coherent supercontinuum generation in a silicon nitride waveguide},}\ }\href
  {\doibase 10.1364/OL.40.005117} {\bibfield  {journal} {\bibinfo  {journal}
  {Optics Letters}\ }\textbf {\bibinfo {volume} {40}},\ \bibinfo {pages}
  {5117--5120} (\bibinfo {year} {2015})}\BibitemShut {NoStop}%
\bibitem [{\citenamefont {Singh}\ \emph {et~al.}(2015)\citenamefont {Singh},
  \citenamefont {Hudson}, \citenamefont {Yu}, \citenamefont {Grillet},
  \citenamefont {Jackson}, \citenamefont {Casas-Bedoya}, \citenamefont {Read},
  \citenamefont {Atanackovic}, \citenamefont {Duvall}, \citenamefont {Palomba},
  \citenamefont {Luther-Davies}, \citenamefont {Madden}, \citenamefont {Moss},\
  and\ \citenamefont {Eggleton}}]{singh_midinfrared_2015}%
  \BibitemOpen
  \bibfield  {author} {\bibinfo {author} {\bibfnamefont {Neetesh}\ \bibnamefont
  {Singh}}, \bibinfo {author} {\bibfnamefont {Darren~D.}\ \bibnamefont
  {Hudson}}, \bibinfo {author} {\bibfnamefont {Yi}~\bibnamefont {Yu}}, \bibinfo
  {author} {\bibfnamefont {Christian}\ \bibnamefont {Grillet}}, \bibinfo
  {author} {\bibfnamefont {Stuart~D.}\ \bibnamefont {Jackson}}, \bibinfo
  {author} {\bibfnamefont {Alvaro}\ \bibnamefont {Casas-Bedoya}}, \bibinfo
  {author} {\bibfnamefont {Andrew}\ \bibnamefont {Read}}, \bibinfo {author}
  {\bibfnamefont {Petar}\ \bibnamefont {Atanackovic}}, \bibinfo {author}
  {\bibfnamefont {Steven~G.}\ \bibnamefont {Duvall}}, \bibinfo {author}
  {\bibfnamefont {Stefano}\ \bibnamefont {Palomba}}, \bibinfo {author}
  {\bibfnamefont {Barry}\ \bibnamefont {Luther-Davies}}, \bibinfo {author}
  {\bibfnamefont {Stephen}\ \bibnamefont {Madden}}, \bibinfo {author}
  {\bibfnamefont {David~J.}\ \bibnamefont {Moss}}, \ and\ \bibinfo {author}
  {\bibfnamefont {Benjamin~J.}\ \bibnamefont {Eggleton}},\ }\bibfield  {title}
  {\enquote {\bibinfo {title} {Midinfrared supercontinuum generation from 2 to
  6 μm in a silicon nanowire},}\ }\href {\doibase 10.1364/OPTICA.2.000797}
  {\bibfield  {journal} {\bibinfo  {journal} {Optica}\ }\textbf {\bibinfo
  {volume} {2}},\ \bibinfo {pages} {797--802} (\bibinfo {year}
  {2015})}\BibitemShut {NoStop}%
\bibitem [{\citenamefont {Hsieh}\ \emph {et~al.}(2007)\citenamefont {Hsieh},
  \citenamefont {Chen}, \citenamefont {Liu}, \citenamefont {Dadap},
  \citenamefont {Panoiu}, \citenamefont {Chou}, \citenamefont {Xia},
  \citenamefont {Green}, \citenamefont {Vlasov},\ and\ \citenamefont
  {Osgood}}]{hsieh_supercontinuum_2007}%
  \BibitemOpen
  \bibfield  {author} {\bibinfo {author} {\bibfnamefont {I.-Wei}\ \bibnamefont
  {Hsieh}}, \bibinfo {author} {\bibfnamefont {Xiaogang}\ \bibnamefont {Chen}},
  \bibinfo {author} {\bibfnamefont {Xiaoping}\ \bibnamefont {Liu}}, \bibinfo
  {author} {\bibfnamefont {Jerry~I.}\ \bibnamefont {Dadap}}, \bibinfo {author}
  {\bibfnamefont {Nicolae~C.}\ \bibnamefont {Panoiu}}, \bibinfo {author}
  {\bibfnamefont {Cheng-Yun}\ \bibnamefont {Chou}}, \bibinfo {author}
  {\bibfnamefont {Fengnian}\ \bibnamefont {Xia}}, \bibinfo {author}
  {\bibfnamefont {William~M.}\ \bibnamefont {Green}}, \bibinfo {author}
  {\bibfnamefont {Yurii~A.}\ \bibnamefont {Vlasov}}, \ and\ \bibinfo {author}
  {\bibfnamefont {Richard~M.}\ \bibnamefont {Osgood}},\ }\bibfield  {title}
  {\enquote {\bibinfo {title} {Supercontinuum generation in silicon photonic
  wires},}\ }\href {\doibase 10.1364/OE.15.015242} {\bibfield  {journal}
  {\bibinfo  {journal} {Optics Express}\ }\textbf {\bibinfo {volume} {15}},\
  \bibinfo {pages} {15242--15249} (\bibinfo {year} {2007})}\BibitemShut
  {NoStop}%
\bibitem [{\citenamefont {Leo}\ \emph {et~al.}(2015)\citenamefont {Leo},
  \citenamefont {Gorza}, \citenamefont {Coen}, \citenamefont {Kuyken},\ and\
  \citenamefont {Roelkens}}]{leo_coherent_2015}%
  \BibitemOpen
  \bibfield  {author} {\bibinfo {author} {\bibfnamefont {François}\
  \bibnamefont {Leo}}, \bibinfo {author} {\bibfnamefont {Simon-Pierre}\
  \bibnamefont {Gorza}}, \bibinfo {author} {\bibfnamefont {Stéphane}\
  \bibnamefont {Coen}}, \bibinfo {author} {\bibfnamefont {Bart}\ \bibnamefont
  {Kuyken}}, \ and\ \bibinfo {author} {\bibfnamefont {Gunther}\ \bibnamefont
  {Roelkens}},\ }\bibfield  {title} {\enquote {\bibinfo {title} {Coherent
  supercontinuum generation in a silicon photonic wire in the telecommunication
  wavelength range},}\ }\href {\doibase 10.1364/OL.40.000123} {\bibfield
  {journal} {\bibinfo  {journal} {Optics Letters}\ }\textbf {\bibinfo {volume}
  {40}},\ \bibinfo {pages} {123--126} (\bibinfo {year} {2015})}\BibitemShut
  {NoStop}%
\bibitem [{\citenamefont {Pu}\ \emph {et~al.}(2016)\citenamefont {Pu},
  \citenamefont {Ji}, \citenamefont {Hu}, \citenamefont {Ottaviano},
  \citenamefont {Semenova}, \citenamefont {Guan}, \citenamefont {Oxenløwe},\
  and\ \citenamefont {Yvind}}]{pu_supercontinuum_2016}%
  \BibitemOpen
  \bibfield  {author} {\bibinfo {author} {\bibfnamefont {Minhao}\ \bibnamefont
  {Pu}}, \bibinfo {author} {\bibfnamefont {Hua}\ \bibnamefont {Ji}}, \bibinfo
  {author} {\bibfnamefont {Hao}\ \bibnamefont {Hu}}, \bibinfo {author}
  {\bibfnamefont {Luisa}\ \bibnamefont {Ottaviano}}, \bibinfo {author}
  {\bibfnamefont {Elizaveta}\ \bibnamefont {Semenova}}, \bibinfo {author}
  {\bibfnamefont {Pengyu}\ \bibnamefont {Guan}}, \bibinfo {author}
  {\bibfnamefont {Leif~K.}\ \bibnamefont {Oxenløwe}}, \ and\ \bibinfo {author}
  {\bibfnamefont {Kresten}\ \bibnamefont {Yvind}},\ }\bibfield  {title}
  {\enquote {\bibinfo {title} {Supercontinuum {Generation} in
  {AlGaAs}-{On}-{Insulator} {Nano}-{Waveguide} at {Telecom} {Wavelengths}},}\
  }in\ \href {\doibase 10.1364/CLEO_AT.2016.AM3J.3} {\emph {\bibinfo
  {booktitle} {Conference on {Lasers} and {Electro}-{Optics} (2016), paper
  {AM}3J.3}}}\ (\bibinfo  {publisher} {Optical Society of America},\ \bibinfo
  {year} {2016})\ p.\ \bibinfo {pages} {AM3J.3}\BibitemShut {NoStop}%
\bibitem [{\citenamefont {Yu}\ \emph {et~al.}(2013)\citenamefont {Yu},
  \citenamefont {Gai}, \citenamefont {Wang}, \citenamefont {Ma}, \citenamefont
  {Wang}, \citenamefont {Yang}, \citenamefont {Choi}, \citenamefont {Madden},\
  and\ \citenamefont {Luther-Davies}}]{yu_mid-infrared_2013}%
  \BibitemOpen
  \bibfield  {author} {\bibinfo {author} {\bibfnamefont {Yi}~\bibnamefont
  {Yu}}, \bibinfo {author} {\bibfnamefont {Xin}\ \bibnamefont {Gai}}, \bibinfo
  {author} {\bibfnamefont {Ting}\ \bibnamefont {Wang}}, \bibinfo {author}
  {\bibfnamefont {Pan}\ \bibnamefont {Ma}}, \bibinfo {author} {\bibfnamefont
  {Rongping}\ \bibnamefont {Wang}}, \bibinfo {author} {\bibfnamefont {Zhiyong}\
  \bibnamefont {Yang}}, \bibinfo {author} {\bibfnamefont {Duk-Yong}\
  \bibnamefont {Choi}}, \bibinfo {author} {\bibfnamefont {Steve}\ \bibnamefont
  {Madden}}, \ and\ \bibinfo {author} {\bibfnamefont {Barry}\ \bibnamefont
  {Luther-Davies}},\ }\bibfield  {title} {\enquote {\bibinfo {title}
  {Mid-infrared supercontinuum generation in chalcogenides},}\ }\href {\doibase
  10.1364/OME.3.001075} {\bibfield  {journal} {\bibinfo  {journal} {Optical
  Materials Express}\ }\textbf {\bibinfo {volume} {3}},\ \bibinfo {pages}
  {1075--1086} (\bibinfo {year} {2013})}\BibitemShut {NoStop}%
\bibitem [{\citenamefont {Lamont}\ \emph {et~al.}(2008)\citenamefont {Lamont},
  \citenamefont {Luther-Davies}, \citenamefont {Choi}, \citenamefont {Madden},\
  and\ \citenamefont {Eggleton}}]{lamont_supercontinuum_2008}%
  \BibitemOpen
  \bibfield  {author} {\bibinfo {author} {\bibfnamefont {Michael R.~E.}\
  \bibnamefont {Lamont}}, \bibinfo {author} {\bibfnamefont {Barry}\
  \bibnamefont {Luther-Davies}}, \bibinfo {author} {\bibfnamefont {Duk-Yong}\
  \bibnamefont {Choi}}, \bibinfo {author} {\bibfnamefont {Steve}\ \bibnamefont
  {Madden}}, \ and\ \bibinfo {author} {\bibfnamefont {Benjamin~J.}\
  \bibnamefont {Eggleton}},\ }\bibfield  {title} {\enquote {\bibinfo {title}
  {Supercontinuum generation in dispersion engineered highly nonlinear (gamma =
  10 /{W}/m) {As}$_{\textrm{2}}${S}$_{\textrm{3}}$ chalcogenide planar
  waveguide},}\ }\href {\doibase 10.1364/OE.16.014938} {\bibfield  {journal}
  {\bibinfo  {journal} {Optics Express}\ }\textbf {\bibinfo {volume} {16}},\
  \bibinfo {pages} {14938--14944} (\bibinfo {year} {2008})}\BibitemShut
  {NoStop}%
\bibitem [{\citenamefont {Iwakuni}\ \emph {et~al.}(2016)\citenamefont
  {Iwakuni}, \citenamefont {Okubo}, \citenamefont {Tadanaga}, \citenamefont
  {Inaba}, \citenamefont {Onae}, \citenamefont {Hong},\ and\ \citenamefont
  {Sasada}}]{iwakuni_generation_2016}%
  \BibitemOpen
  \bibfield  {author} {\bibinfo {author} {\bibfnamefont {Kana}\ \bibnamefont
  {Iwakuni}}, \bibinfo {author} {\bibfnamefont {Sho}\ \bibnamefont {Okubo}},
  \bibinfo {author} {\bibfnamefont {Osamu}\ \bibnamefont {Tadanaga}}, \bibinfo
  {author} {\bibfnamefont {Hajime}\ \bibnamefont {Inaba}}, \bibinfo {author}
  {\bibfnamefont {Atsushi}\ \bibnamefont {Onae}}, \bibinfo {author}
  {\bibfnamefont {Feng-Lei}\ \bibnamefont {Hong}}, \ and\ \bibinfo {author}
  {\bibfnamefont {Hiroyuki}\ \bibnamefont {Sasada}},\ }\bibfield  {title}
  {\enquote {\bibinfo {title} {Generation of a frequency comb spanning more
  than 3.6 octaves from ultraviolet to mid infrared},}\ }\href {\doibase
  10.1364/OL.41.003980} {\bibfield  {journal} {\bibinfo  {journal} {Optics
  Letters}\ }\textbf {\bibinfo {volume} {41}},\ \bibinfo {pages} {3980--3983}
  (\bibinfo {year} {2016})}\BibitemShut {NoStop}%
\bibitem [{\citenamefont {Guo}\ \emph {et~al.}(2015)\citenamefont {Guo},
  \citenamefont {Zhou}, \citenamefont {Steinert}, \citenamefont {Setzpfandt},
  \citenamefont {Pertsch}, \citenamefont {Chung}, \citenamefont {Chen},\ and\
  \citenamefont {Bache}}]{guo_supercontinuum_2015}%
  \BibitemOpen
  \bibfield  {author} {\bibinfo {author} {\bibfnamefont {Hairun}\ \bibnamefont
  {Guo}}, \bibinfo {author} {\bibfnamefont {Binbin}\ \bibnamefont {Zhou}},
  \bibinfo {author} {\bibfnamefont {Michael}\ \bibnamefont {Steinert}},
  \bibinfo {author} {\bibfnamefont {Frank}\ \bibnamefont {Setzpfandt}},
  \bibinfo {author} {\bibfnamefont {Thomas}\ \bibnamefont {Pertsch}}, \bibinfo
  {author} {\bibfnamefont {Hung-ping}\ \bibnamefont {Chung}}, \bibinfo {author}
  {\bibfnamefont {Yen-Hung}\ \bibnamefont {Chen}}, \ and\ \bibinfo {author}
  {\bibfnamefont {Morten}\ \bibnamefont {Bache}},\ }\bibfield  {title}
  {\enquote {\bibinfo {title} {Supercontinuum generation in quadratic nonlinear
  waveguides without quasi-phase matching},}\ }\href {\doibase
  10.1364/OL.40.000629} {\bibfield  {journal} {\bibinfo  {journal} {Optics
  Letters}\ }\textbf {\bibinfo {volume} {40}},\ \bibinfo {pages} {629--632}
  (\bibinfo {year} {2015})}\BibitemShut {NoStop}%
\bibitem [{\citenamefont {Langrock}\ \emph {et~al.}(2007)\citenamefont
  {Langrock}, \citenamefont {Fejer}, \citenamefont {Hartl},\ and\ \citenamefont
  {Fermann}}]{langrock_generation_2007}%
  \BibitemOpen
  \bibfield  {author} {\bibinfo {author} {\bibfnamefont {Carsten}\ \bibnamefont
  {Langrock}}, \bibinfo {author} {\bibfnamefont {M.~M.}\ \bibnamefont {Fejer}},
  \bibinfo {author} {\bibfnamefont {I.}~\bibnamefont {Hartl}}, \ and\ \bibinfo
  {author} {\bibfnamefont {Martin~E.}\ \bibnamefont {Fermann}},\ }\bibfield
  {title} {\enquote {\bibinfo {title} {Generation of octave-spanning spectra
  inside reverse-proton-exchanged periodically poled lithium niobate
  waveguides},}\ }\href {\doibase 10.1364/OL.32.002478} {\bibfield  {journal}
  {\bibinfo  {journal} {Optics Letters}\ }\textbf {\bibinfo {volume} {32}},\
  \bibinfo {pages} {2478--2480} (\bibinfo {year} {2007})}\BibitemShut {NoStop}%
\bibitem [{\citenamefont {Guo}\ \emph {et~al.}(2016{\natexlab{a}})\citenamefont
  {Guo}, \citenamefont {Zou},\ and\ \citenamefont
  {Tang}}]{guo_second-harmonic_2016}%
  \BibitemOpen
  \bibfield  {author} {\bibinfo {author} {\bibfnamefont {Xiang}\ \bibnamefont
  {Guo}}, \bibinfo {author} {\bibfnamefont {Chang-Ling}\ \bibnamefont {Zou}}, \
  and\ \bibinfo {author} {\bibfnamefont {Hong~X.}\ \bibnamefont {Tang}},\
  }\bibfield  {title} {\enquote {\bibinfo {title} {Second-harmonic generation
  in aluminum nitride microrings with 2500\%/{W} conversion efficiency},}\
  }\href {\doibase 10.1364/OPTICA.3.001126} {\bibfield  {journal} {\bibinfo
  {journal} {Optica}\ }\textbf {\bibinfo {volume} {3}},\ \bibinfo {pages}
  {1126--1131} (\bibinfo {year} {2016}{\natexlab{a}})}\BibitemShut {NoStop}%
\bibitem [{\citenamefont {Jung}\ \emph {et~al.}(2013)\citenamefont {Jung},
  \citenamefont {Xiong}, \citenamefont {Fong}, \citenamefont {Zhang},\ and\
  \citenamefont {Tang}}]{jung_optical_2013}%
  \BibitemOpen
  \bibfield  {author} {\bibinfo {author} {\bibfnamefont {Hojoong}\ \bibnamefont
  {Jung}}, \bibinfo {author} {\bibfnamefont {Chi}\ \bibnamefont {Xiong}},
  \bibinfo {author} {\bibfnamefont {King~Y.}\ \bibnamefont {Fong}}, \bibinfo
  {author} {\bibfnamefont {Xufeng}\ \bibnamefont {Zhang}}, \ and\ \bibinfo
  {author} {\bibfnamefont {Hong~X.}\ \bibnamefont {Tang}},\ }\bibfield  {title}
  {\enquote {\bibinfo {title} {Optical frequency comb generation from aluminum
  nitride microring resonator},}\ }\href {\doibase 10.1364/OL.38.002810}
  {\bibfield  {journal} {\bibinfo  {journal} {Optics Letters}\ }\textbf
  {\bibinfo {volume} {38}},\ \bibinfo {pages} {2810--2813} (\bibinfo {year}
  {2013})}\BibitemShut {NoStop}%
\bibitem [{\citenamefont {Zhao}\ \emph {et~al.}(2015)\citenamefont {Zhao},
  \citenamefont {Connie}, \citenamefont {Dastjerdi}, \citenamefont {Kong},
  \citenamefont {Wang}, \citenamefont {Djavid}, \citenamefont {Sadaf},
  \citenamefont {Liu}, \citenamefont {Shih}, \citenamefont {Guo},\ and\
  \citenamefont {Mi}}]{zhao_aluminum_2015}%
  \BibitemOpen
  \bibfield  {author} {\bibinfo {author} {\bibfnamefont {S.}~\bibnamefont
  {Zhao}}, \bibinfo {author} {\bibfnamefont {A.~T.}\ \bibnamefont {Connie}},
  \bibinfo {author} {\bibfnamefont {M.~H.~T.}\ \bibnamefont {Dastjerdi}},
  \bibinfo {author} {\bibfnamefont {X.~H.}\ \bibnamefont {Kong}}, \bibinfo
  {author} {\bibfnamefont {Q.}~\bibnamefont {Wang}}, \bibinfo {author}
  {\bibfnamefont {M.}~\bibnamefont {Djavid}}, \bibinfo {author} {\bibfnamefont
  {S.}~\bibnamefont {Sadaf}}, \bibinfo {author} {\bibfnamefont {X.~D.}\
  \bibnamefont {Liu}}, \bibinfo {author} {\bibfnamefont {I.}~\bibnamefont
  {Shih}}, \bibinfo {author} {\bibfnamefont {H.}~\bibnamefont {Guo}}, \ and\
  \bibinfo {author} {\bibfnamefont {Z.}~\bibnamefont {Mi}},\ }\bibfield
  {title} {\enquote {\bibinfo {title} {Aluminum nitride nanowire light emitting
  diodes: {Breaking} the fundamental bottleneck of deep ultraviolet light
  sources},}\ }\href {\doibase 10.1038/srep08332} {\bibfield  {journal}
  {\bibinfo  {journal} {Scientific Reports}\ }\textbf {\bibinfo {volume} {5}},\
  \bibinfo {pages} {8332} (\bibinfo {year} {2015})}\BibitemShut {NoStop}%
\bibitem [{\citenamefont {Xiong}\ \emph {et~al.}(2012)\citenamefont {Xiong},
  \citenamefont {Pernice},\ and\ \citenamefont {Tang}}]{xiong_low-loss_2012}%
  \BibitemOpen
  \bibfield  {author} {\bibinfo {author} {\bibfnamefont {Chi}\ \bibnamefont
  {Xiong}}, \bibinfo {author} {\bibfnamefont {Wolfram H.~P.}\ \bibnamefont
  {Pernice}}, \ and\ \bibinfo {author} {\bibfnamefont {Hong~X.}\ \bibnamefont
  {Tang}},\ }\bibfield  {title} {\enquote {\bibinfo {title} {Low-{Loss},
  {Silicon} {Integrated}, {Aluminum} {Nitride} {Photonic} {Circuits} and
  {Their} {Use} for {Electro}-{Optic} {Signal} {Processing}},}\ }\href
  {\doibase 10.1021/nl3011885} {\bibfield  {journal} {\bibinfo  {journal} {Nano
  Letters}\ }\textbf {\bibinfo {volume} {12}},\ \bibinfo {pages} {3562--3568}
  (\bibinfo {year} {2012})}\BibitemShut {NoStop}%
\bibitem [{\citenamefont {Sinclair}\ \emph {et~al.}(2015)\citenamefont
  {Sinclair}, \citenamefont {Deschênes}, \citenamefont {Sonderhouse},
  \citenamefont {Swann}, \citenamefont {Khader}, \citenamefont {Baumann},
  \citenamefont {Newbury},\ and\ \citenamefont
  {Coddington}}]{sinclair_compact_2015}%
  \BibitemOpen
  \bibfield  {author} {\bibinfo {author} {\bibfnamefont {L.~C.}\ \bibnamefont
  {Sinclair}}, \bibinfo {author} {\bibfnamefont {J.-D.}\ \bibnamefont
  {Deschênes}}, \bibinfo {author} {\bibfnamefont {L.}~\bibnamefont
  {Sonderhouse}}, \bibinfo {author} {\bibfnamefont {W.~C.}\ \bibnamefont
  {Swann}}, \bibinfo {author} {\bibfnamefont {I.~H.}\ \bibnamefont {Khader}},
  \bibinfo {author} {\bibfnamefont {E.}~\bibnamefont {Baumann}}, \bibinfo
  {author} {\bibfnamefont {N.~R.}\ \bibnamefont {Newbury}}, \ and\ \bibinfo
  {author} {\bibfnamefont {I.}~\bibnamefont {Coddington}},\ }\bibfield  {title}
  {\enquote {\bibinfo {title} {A compact optically coherent fiber frequency
  comb},}\ }\href {\doibase 10.1063/1.4928163} {\bibfield  {journal} {\bibinfo
  {journal} {Review of Scientific Instruments}\ }\textbf {\bibinfo {volume}
  {86}},\ \bibinfo {pages} {081301} (\bibinfo {year} {2015})}\BibitemShut
  {NoStop}%
\bibitem [{\citenamefont {Hult}(2007)}]{hult_fourth-order_2007}%
  \BibitemOpen
  \bibfield  {author} {\bibinfo {author} {\bibfnamefont {J.}~\bibnamefont
  {Hult}},\ }\bibfield  {title} {\enquote {\bibinfo {title} {A {Fourth}-{Order}
  {Runge}-{Kutta} in the {Interaction} {Picture} {Method} for {Simulating}
  {Supercontinuum} {Generation} in {Optical} {Fibers}},}\ }\href {\doibase
  10.1109/JLT.2007.909373} {\bibfield  {journal} {\bibinfo  {journal} {Journal
  of Lightwave Technology}\ }\textbf {\bibinfo {volume} {25}},\ \bibinfo
  {pages} {3770--3775} (\bibinfo {year} {2007})}\BibitemShut {NoStop}%
\bibitem [{\citenamefont {Heidt}(2009)}]{heidt_efficient_2009}%
  \BibitemOpen
  \bibfield  {author} {\bibinfo {author} {\bibfnamefont {A.~M.}\ \bibnamefont
  {Heidt}},\ }\bibfield  {title} {\enquote {\bibinfo {title} {Efficient
  {Adaptive} {Step} {Size} {Method} for the {Simulation} of {Supercontinuum}
  {Generation} in {Optical} {Fibers}},}\ }\href {\doibase
  10.1109/JLT.2009.2021538} {\bibfield  {journal} {\bibinfo  {journal} {Journal
  of Lightwave Technology}\ }\textbf {\bibinfo {volume} {27}},\ \bibinfo
  {pages} {3984--3991} (\bibinfo {year} {2009})}\BibitemShut {NoStop}%
\bibitem [{\citenamefont {Ycas}\ \emph {et~al.}(2016)\citenamefont {Ycas},
  \citenamefont {Maser},\ and\ \citenamefont {Hickstein}}]{ycas_pynlo_2016}%
  \BibitemOpen
  \bibfield  {author} {\bibinfo {author} {\bibfnamefont {Gabriel}\ \bibnamefont
  {Ycas}}, \bibinfo {author} {\bibfnamefont {Daniel}\ \bibnamefont {Maser}}, \
  and\ \bibinfo {author} {\bibfnamefont {Daniel~D.}\ \bibnamefont
  {Hickstein}},\ }\href {https://github.com/pyNLO/PyNLO} {\enquote {\bibinfo
  {title} {{pyNLO} - {Nonlinear} optics modeling for {Python}},}\ } (\bibinfo
  {year} {2016})\BibitemShut {NoStop}%
\bibitem [{\citenamefont {Amorim}\ \emph {et~al.}(2009)\citenamefont {Amorim},
  \citenamefont {Tognetti}, \citenamefont {Oliveira}, \citenamefont {Silva},
  \citenamefont {Bernardo}, \citenamefont {Kärtner},\ and\ \citenamefont
  {Crespo}}]{amorim_sub_2009}%
  \BibitemOpen
  \bibfield  {author} {\bibinfo {author} {\bibfnamefont {A.~A.}\ \bibnamefont
  {Amorim}}, \bibinfo {author} {\bibfnamefont {M.~V.}\ \bibnamefont
  {Tognetti}}, \bibinfo {author} {\bibfnamefont {P.}~\bibnamefont {Oliveira}},
  \bibinfo {author} {\bibfnamefont {J.~L.}\ \bibnamefont {Silva}}, \bibinfo
  {author} {\bibfnamefont {L.~M.}\ \bibnamefont {Bernardo}}, \bibinfo {author}
  {\bibfnamefont {F.~X.}\ \bibnamefont {Kärtner}}, \ and\ \bibinfo {author}
  {\bibfnamefont {H.~M.}\ \bibnamefont {Crespo}},\ }\bibfield  {title}
  {\enquote {\bibinfo {title} {Sub two cycle pulses by soliton self-compression
  in highly nonlinear photonic crystal fibers},}\ }\href {\doibase
  10.1364/OL.34.003851} {\bibfield  {journal} {\bibinfo  {journal} {Optics
  Letters}\ }\textbf {\bibinfo {volume} {34}},\ \bibinfo {pages} {3851--3853}
  (\bibinfo {year} {2009})}\BibitemShut {NoStop}%
\bibitem [{\citenamefont {Fallahkhair}\ \emph {et~al.}(2008)\citenamefont
  {Fallahkhair}, \citenamefont {Li},\ and\ \citenamefont
  {Murphy}}]{fallahkhair_vector_2008}%
  \BibitemOpen
  \bibfield  {author} {\bibinfo {author} {\bibfnamefont {A.~B.}\ \bibnamefont
  {Fallahkhair}}, \bibinfo {author} {\bibfnamefont {K.~S.}\ \bibnamefont {Li}},
  \ and\ \bibinfo {author} {\bibfnamefont {T.~E.}\ \bibnamefont {Murphy}},\
  }\bibfield  {title} {\enquote {\bibinfo {title} {Vector {Finite} {Difference}
  {Modesolver} for {Anisotropic} {Dielectric} {Waveguides}},}\ }\href {\doibase
  10.1109/JLT.2008.923643} {\bibfield  {journal} {\bibinfo  {journal} {Journal
  of Lightwave Technology}\ }\textbf {\bibinfo {volume} {26}},\ \bibinfo
  {pages} {1423--1431} (\bibinfo {year} {2008})}\BibitemShut {NoStop}%
\bibitem [{\citenamefont {Akhmediev}\ and\ \citenamefont
  {Karlsson}(1995)}]{akhmediev_cherenkov_1995}%
  \BibitemOpen
  \bibfield  {author} {\bibinfo {author} {\bibfnamefont {Nail}\ \bibnamefont
  {Akhmediev}}\ and\ \bibinfo {author} {\bibfnamefont {Magnus}\ \bibnamefont
  {Karlsson}},\ }\bibfield  {title} {\enquote {\bibinfo {title} {Cherenkov
  radiation emitted by solitons in optical fibers},}\ }\href {\doibase
  10.1103/PhysRevA.51.2602} {\bibfield  {journal} {\bibinfo  {journal}
  {Physical Review A}\ }\textbf {\bibinfo {volume} {51}},\ \bibinfo {pages}
  {2602--2607} (\bibinfo {year} {1995})}\BibitemShut {NoStop}%
\bibitem [{\citenamefont {Solomons}\ and\ \citenamefont
  {Fryhle}(2009)}]{solomons_organic_2009}%
  \BibitemOpen
  \bibfield  {author} {\bibinfo {author} {\bibfnamefont {T.~W.~Graham}\
  \bibnamefont {Solomons}}\ and\ \bibinfo {author} {\bibfnamefont {Craig~B.}\
  \bibnamefont {Fryhle}},\ }\href@noop {} {{\selectlanguage {English}\emph
  {\bibinfo {title} {Organic {Chemistry}}}}},\ \bibinfo {edition} {10th}\ ed.\
  (\bibinfo  {publisher} {Wiley},\ \bibinfo {year} {2009})\BibitemShut
  {NoStop}%
\bibitem [{\citenamefont {Navarra}\ \emph {et~al.}(2005)\citenamefont
  {Navarra}, \citenamefont {Iliopoulos}, \citenamefont {Militello},
  \citenamefont {Rotolo},\ and\ \citenamefont {Leone}}]{navarra_oh_2005}%
  \BibitemOpen
  \bibfield  {author} {\bibinfo {author} {\bibfnamefont {G.}~\bibnamefont
  {Navarra}}, \bibinfo {author} {\bibfnamefont {I.}~\bibnamefont {Iliopoulos}},
  \bibinfo {author} {\bibfnamefont {V.}~\bibnamefont {Militello}}, \bibinfo
  {author} {\bibfnamefont {S.~G.}\ \bibnamefont {Rotolo}}, \ and\ \bibinfo
  {author} {\bibfnamefont {M.}~\bibnamefont {Leone}},\ }\bibfield  {title}
  {\enquote {\bibinfo {title} {{OH} related infrared absorption bands in oxide
  glasses},}\ }\href {\doibase 10.1016/j.jnoncrysol.2005.04.018} {\bibfield
  {journal} {\bibinfo  {journal} {Journal of Non-Crystalline Solids}\ }\textbf
  {\bibinfo {volume} {351}},\ \bibinfo {pages} {1796--1800} (\bibinfo {year}
  {2005})}\BibitemShut {NoStop}%
\bibitem [{\citenamefont {Cole}\ \emph {et~al.}(2016)\citenamefont {Cole},
  \citenamefont {Lamb}, \citenamefont {Del'Haye}, \citenamefont {Diddams},\
  and\ \citenamefont {Papp}}]{cole_soliton_2016}%
  \BibitemOpen
  \bibfield  {author} {\bibinfo {author} {\bibfnamefont {Daniel~C.}\
  \bibnamefont {Cole}}, \bibinfo {author} {\bibfnamefont {Erin~S.}\
  \bibnamefont {Lamb}}, \bibinfo {author} {\bibfnamefont {Pascal}\ \bibnamefont
  {Del'Haye}}, \bibinfo {author} {\bibfnamefont {Scott~A.}\ \bibnamefont
  {Diddams}}, \ and\ \bibinfo {author} {\bibfnamefont {Scott~B.}\ \bibnamefont
  {Papp}},\ }\bibfield  {title} {\enquote {\bibinfo {title} {Soliton crystals
  in {Kerr} resonators},}\ }\href {http://arxiv.org/abs/1610.00080} {\bibfield
  {journal} {\bibinfo  {journal} {arXiv:1610.00080 [physics]}\ } (\bibinfo
  {year} {2016})},\ \bibinfo {note} {arXiv: 1610.00080}\BibitemShut {NoStop}%
\bibitem [{\citenamefont {Ramelow}\ \emph {et~al.}(2014)\citenamefont
  {Ramelow}, \citenamefont {Farsi}, \citenamefont {Clemmen}, \citenamefont
  {Levy}, \citenamefont {Johnson}, \citenamefont {Okawachi}, \citenamefont
  {Lamont}, \citenamefont {Lipson},\ and\ \citenamefont
  {Gaeta}}]{ramelow_strong_2014}%
  \BibitemOpen
  \bibfield  {author} {\bibinfo {author} {\bibfnamefont {Sven}\ \bibnamefont
  {Ramelow}}, \bibinfo {author} {\bibfnamefont {Alessandro}\ \bibnamefont
  {Farsi}}, \bibinfo {author} {\bibfnamefont {Stéphane}\ \bibnamefont
  {Clemmen}}, \bibinfo {author} {\bibfnamefont {Jacob~S.}\ \bibnamefont
  {Levy}}, \bibinfo {author} {\bibfnamefont {Adrea~R.}\ \bibnamefont
  {Johnson}}, \bibinfo {author} {\bibfnamefont {Yoshitomo}\ \bibnamefont
  {Okawachi}}, \bibinfo {author} {\bibfnamefont {Michael R.~E.}\ \bibnamefont
  {Lamont}}, \bibinfo {author} {\bibfnamefont {Michal}\ \bibnamefont {Lipson}},
  \ and\ \bibinfo {author} {\bibfnamefont {Alexander~L.}\ \bibnamefont
  {Gaeta}},\ }\bibfield  {title} {\enquote {\bibinfo {title} {Strong
  polarization mode coupling in microresonators},}\ }\href {\doibase
  10.1364/OL.39.005134} {\bibfield  {journal} {\bibinfo  {journal} {Optics
  Letters}\ }\textbf {\bibinfo {volume} {39}},\ \bibinfo {pages} {5134--5137}
  (\bibinfo {year} {2014})}\BibitemShut {NoStop}%
\bibitem [{\citenamefont {Herr}\ \emph
  {et~al.}(2014{\natexlab{b}})\citenamefont {Herr}, \citenamefont {Brasch},
  \citenamefont {Jost}, \citenamefont {Mirgorodskiy}, \citenamefont {Lihachev},
  \citenamefont {Gorodetsky},\ and\ \citenamefont
  {Kippenberg}}]{herr_mode_2014}%
  \BibitemOpen
  \bibfield  {author} {\bibinfo {author} {\bibfnamefont {T.}~\bibnamefont
  {Herr}}, \bibinfo {author} {\bibfnamefont {V.}~\bibnamefont {Brasch}},
  \bibinfo {author} {\bibfnamefont {J. D.}\ \bibnamefont {Jost}}, \bibinfo
  {author} {\bibfnamefont {I.}~\bibnamefont {Mirgorodskiy}}, \bibinfo {author}
  {\bibfnamefont {G.}~\bibnamefont {Lihachev}}, \bibinfo {author}
  {\bibfnamefont {M. L.}\ \bibnamefont {Gorodetsky}}, \ and\ \bibinfo
  {author} {\bibfnamefont {T. J.}\ \bibnamefont {Kippenberg}},\ }\bibfield
  {title} {\enquote {\bibinfo {title} {Mode {Spectrum} and {Temporal} {Soliton}
  {Formation} in {Optical} {Microresonators}},}\ }\href {\doibase
  10.1103/PhysRevLett.113.123901} {\bibfield  {journal} {\bibinfo  {journal}
  {Physical Review Letters}\ }\textbf {\bibinfo {volume} {113}},\ \bibinfo
  {pages} {123901} (\bibinfo {year} {2014}{\natexlab{b}})}\BibitemShut
  {NoStop}%
\bibitem [{\citenamefont {Carlson}\ \emph
  {et~al.}(2017{\natexlab{b}})\citenamefont {Carlson}, \citenamefont
  {Hickstein}, \citenamefont {Lind}, \citenamefont {Droste}, \citenamefont
  {Westly}, \citenamefont {Nader}, \citenamefont {Coddington}, \citenamefont
  {Newbury}, \citenamefont {Srinivasan}, \citenamefont {Diddams},\ and\
  \citenamefont {Papp}}]{carlson_high-efficiency_2017}%
  \BibitemOpen
  \bibfield  {author} {\bibinfo {author} {\bibfnamefont {David}\ \bibnamefont
  {Carlson}}, \bibinfo {author} {\bibfnamefont {Daniel~D.}\ \bibnamefont
  {Hickstein}}, \bibinfo {author} {\bibfnamefont {Alex}\ \bibnamefont {Lind}},
  \bibinfo {author} {\bibfnamefont {Stefan}\ \bibnamefont {Droste}}, \bibinfo
  {author} {\bibfnamefont {Daron}\ \bibnamefont {Westly}}, \bibinfo {author}
  {\bibfnamefont {Nima}\ \bibnamefont {Nader}}, \bibinfo {author}
  {\bibfnamefont {Ian~R.}\ \bibnamefont {Coddington}}, \bibinfo {author}
  {\bibfnamefont {Nathan~R.}\ \bibnamefont {Newbury}}, \bibinfo {author}
  {\bibfnamefont {Kartik}\ \bibnamefont {Srinivasan}}, \bibinfo {author}
  {\bibfnamefont {Scott~A.}\ \bibnamefont {Diddams}}, \ and\ \bibinfo {author}
  {\bibfnamefont {Scott~B.}\ \bibnamefont {Papp}},\ }\bibfield  {title}
  {\enquote {\bibinfo {title} {High-efficiency wavelength conversion in silicon
  nitride waveguides for self-referenced frequency combs},}\ }\href@noop {}
  {\bibfield  {journal} {\bibinfo  {journal} {In preparation}\ } (\bibinfo
  {year} {2017}{\natexlab{b}})}\BibitemShut {NoStop}%
\bibitem [{\citenamefont {Guo}\ \emph {et~al.}(2016{\natexlab{b}})\citenamefont
  {Guo}, \citenamefont {Zou}, \citenamefont {Jung},\ and\ \citenamefont
  {Tang}}]{guo_chip_2016}%
  \BibitemOpen
  \bibfield  {author} {\bibinfo {author} {\bibfnamefont {Xiang}\ \bibnamefont
  {Guo}}, \bibinfo {author} {\bibfnamefont {Chang-Ling}\ \bibnamefont {Zou}},
  \bibinfo {author} {\bibfnamefont {Hojoong}\ \bibnamefont {Jung}}, \ and\
  \bibinfo {author} {\bibfnamefont {Hong~X.}\ \bibnamefont {Tang}},\ }\bibfield
   {title} {\enquote {\bibinfo {title} {On {Chip} {Strong} {Coupling} and
  {Efficient} {Frequency} {Conversion} between {Telecom} and {Visible}
  {Optical} {Modes}},}\ }\href {\doibase 10.1103/PhysRevLett.117.123902}
  {\bibfield  {journal} {\bibinfo  {journal} {Physical Review Letters}\
  }\textbf {\bibinfo {volume} {117}},\ \bibinfo {pages} {123902} (\bibinfo
  {year} {2016}{\natexlab{b}})}\BibitemShut {NoStop}%
\bibitem [{\citenamefont {Dawkins}\ \emph {et~al.}(2007)\citenamefont
  {Dawkins}, \citenamefont {McFerran},\ and\ \citenamefont
  {Luiten}}]{dawkins_considerations_2007}%
  \BibitemOpen
  \bibfield  {author} {\bibinfo {author} {\bibfnamefont {S.~T.}\ \bibnamefont
  {Dawkins}}, \bibinfo {author} {\bibfnamefont {J.~J.}\ \bibnamefont
  {McFerran}}, \ and\ \bibinfo {author} {\bibfnamefont {A.~N.}\ \bibnamefont
  {Luiten}},\ }\bibfield  {title} {\enquote {\bibinfo {title} {Considerations
  on the measurement of the stability of oscillators with frequency
  counters},}\ }\href {\doibase 10.1109/TUFFC.2007.337} {\bibfield  {journal}
  {\bibinfo  {journal} {IEEE Transactions on Ultrasonics, Ferroelectrics, and
  Frequency Control}\ }\textbf {\bibinfo {volume} {54}},\ \bibinfo {pages}
  {918--925} (\bibinfo {year} {2007})}\BibitemShut {NoStop}%
\bibitem [{\citenamefont {Jung}\ \emph {et~al.}(2016)\citenamefont {Jung},
  \citenamefont {Guo}, \citenamefont {Zhu}, \citenamefont {Papp}, \citenamefont
  {Diddams},\ and\ \citenamefont {Tang}}]{jung_phase-dependent_2016}%
  \BibitemOpen
  \bibfield  {author} {\bibinfo {author} {\bibfnamefont {Hojoong}\ \bibnamefont
  {Jung}}, \bibinfo {author} {\bibfnamefont {Xiang}\ \bibnamefont {Guo}},
  \bibinfo {author} {\bibfnamefont {Na}~\bibnamefont {Zhu}}, \bibinfo {author}
  {\bibfnamefont {Scott~B.}\ \bibnamefont {Papp}}, \bibinfo {author}
  {\bibfnamefont {Scott~A.}\ \bibnamefont {Diddams}}, \ and\ \bibinfo {author}
  {\bibfnamefont {Hong~X.}\ \bibnamefont {Tang}},\ }\bibfield  {title}
  {\enquote {\bibinfo {title} {Phase-dependent interference between frequency
  doubled comb lines in a {X}$^{\textrm{(2)}}$ phase-matched aluminum nitride
  microring},}\ }\href {\doibase 10.1364/OL.41.003747} {\bibfield  {journal}
  {\bibinfo  {journal} {Optics Letters}\ }\textbf {\bibinfo {volume} {41}},\
  \bibinfo {pages} {3747--3750} (\bibinfo {year} {2016})}\BibitemShut {NoStop}%
\end{thebibliography}%

\end{document}